\documentclass[%
 reprint,
superscriptaddress,
 amsmath,amssymb,
 aps,
pra,
floatfix,
]{revtex4-2}
\usepackage{color}
\usepackage{graphicx}
\usepackage{dcolumn}
\usepackage{bm}
\usepackage[hidelinks]{hyperref}
\usepackage{upgreek}
\usepackage{float}
\usepackage{lineno}

\begin{document}

\title{Cascaded Multiparameter Quantum Metrology}

\preprint{APS/123-QED}

\author{Gregory Krueper}
\affiliation{Department of Physics, University of Colorado Boulder, Boulder CO 80309, USA}
\author{Lior Cohen}
\affiliation{Department of Electrical, Computer and Energy Engineering, University of Colorado Boulder, Boulder CO 80309, USA} 
\author{Juliet T. Gopinath}
\affiliation{Department of Physics, University of Colorado Boulder, Boulder CO 80309, USA} 
\affiliation{Department of Electrical, Computer and Energy Engineering, University of Colorado Boulder, Boulder CO 80309, USA}
\affiliation{Materials Science and Engineering Program, University of Colorado Boulder, Boulder CO 80309, USA}
\email{julietg@colorado.edu}


\begin{abstract}
We present an innovative, platform-independent concept for multiparameter sensing where the measurable parameters are in series, or cascaded, enabling measurements as a function of position. 
With temporally resolved detection, we show that squeezing can give a quantum enhancement in sensitivity over that of classical states by a factor of $e^{2r}$, where $r \approx 1$ is the squeezing parameter. 
As an example, we have modeled an interferometer that senses multiple phase shifts along the same path, demonstrating a maximal quantum advantage by combining a coherent state with squeezed vacuum.
Further classical modeling with up to 100 phases shows linear scaling potential for adding nodes to the sensor. 
The approach can be applied to remote sensing, geophysical surveying, and infrastructure monitoring.
\end{abstract}
\maketitle
\section{Introduction}
Precision measurement from a sensor is often limited by an intrinsic uncertainty from quantum fluctuations. 
Non-classical resources such as entanglement or squeezing can reduce these uncertainties for higher sensitivity \cite{Caves1981, Giovannetti2004, Dowling2015, Bondurant1984, Aasi2013, Taylor2013}. 
This enhancement can be modeled in the framework of quantum metrology, which finds the optimal quantum state and measurement for estimating a given parameter. 
A common quantum state to consider is the squeezed state, which has non-Poissonian photon statistics that give lower measurement uncertainty in one of its quadrature operators \cite{Caves1981}.
Measurements beyond the classical limit were shown in both atomic systems with spin squeezing \cite{MA201189} and in optical systems with electric field quadrature squeezing \cite{Caves1981, Aasi2013}. With state-of-the-art squeezed sources, the sensitivity enhancement can be around an order of magnitude \cite{Mehmet2011, Cox2016}.

While quantum-enhanced sensing is well understood for estimating a single parameter, simultaneous estimation of multiple parameters is far more complex \cite{Crowley2014, Genoni2011, Vidrighin2014}. 
Many sensing problems are inherently multidimensional, such as magnetic or gravitational field sensing \cite{Li2018, Friel2020, Hou2020}, process tomography \cite{Komar2014, Suzuki2020}, localization microscopy \cite{Bisketzi2019}, and optical imaging \cite{Albarelli2020, Fiderer2021}. 
Many studies analyze estimation of different types of parameters simultaneously \cite{Ragy2016, Szczykulska2016, Albarelli2020, Fiderer2021, Albarelli2022}, such as both a phase shift and optical loss \cite{Crowley2014}, or both phase and phase diffusion \cite{Vidrighin2014}.
Others focus on massively parallelized estimation of parameters, all of the same type \cite{Gagatsos2016, You2017, Gorecki2022}.
Since a real sensor will estimate any number of parameters in any configuration, it is critical to develop multiparameter sensing to handle the many variants and geometries of precision sensors.

An established example of multiparameter estimation is phase imaging, which splits an input state across many parallel spatial modes, as shown in Fig. \ref{fig:concept}(a) \cite{Ragy2016,Humphreys2013, Gagatsos2016, Gorecki2022, Hou2020, Polino2019, You2017, Zhang2017}. 
Each mode modifies the state with phase $\phi_i$.
After a mixing circuit interferes these modes together, a joint multimode measurement estimates each parameter.
There are two sensitivity enhancements to consider.
The first is an enhancement from a joint measurement of $N$ parameters, when compared to $N$ separate measurements (advantage \emph{A}). 
The second is an enhancement from using quantum states and measurements, when compared to using classical methods (advantage \emph{B}).
For advantage \emph{A}, without N-mode entanglement, the parallelized multiparameter sensor shows a constant-factor improvement because the reference arm uses fewer resources \cite{Gagatsos2016, You2017, Gorecki2022}. It is predicted that $N$-mode entanglement can yield a factor of $N$ improvement in variance \cite{Humphreys2013}, but this method seems to require contradictory assumptions for saturating the quantum Fisher information \cite{Gorecki2022}. The factor of $N$ advantage is more easily demonstrated when the architecture instead functions as a distributed sensor, which estimates a linear combination of parameters \cite{Malitesta2023, Valeri2023, Proctor2018, Oh2020, Guo2020, Gatto2019, Ge2018}.

For advantage \emph{B}, the Heisenberg limit in sensitivity implies that the measurement variance on any one parameter can at most scale as $\bar{n}^{-2}$, where $\bar{n}$ is the average photon number of the state. Heisenberg scaling is a factor of $\bar{n}$ better than what is achievable with classical resources \cite{Dowling2015}. With entangled states, the enhancement is achieved by the state changing $\bar{n}$ times faster than a classical state of the same energy. Alternatively, squeezed states can reduce the inherent variance in a measured quadrature by a factor of $e^{2r}$ (at the cost of larger noise in the orthogonal, antisqueezed quadrature) \cite{Caves1981}. More specifically, with a squeezed vacuum + coherent state used in LIGO, the squeezing parameter $r$ sets the maximum quantum advantage in variance to $e^{2r}$.
These scaling terms remain when moving to multiparameter estimation \cite{Humphreys2013, Gagatsos2016}.
The parallelized architecture demonstrates both advantage \emph{A} and advantage \emph{B} because each parameter is isolated to a unique mode so that the parameters are individually addressable.
Consequently, a single measurement can optimally estimate all parameters.

\begin{figure}[t]
    \centering
    \includegraphics[keepaspectratio, width=0.8\columnwidth]{concept_fig.png}
    \caption{(a) Conceptual layout of a parallelized multiparameter sensor. A probe is split into several parallel modes, each with one parameter. If multiple parameters lie on the same mode, like $\phi_4$ and $\phi_5$, then the detectors are unable to distinguish between them. Instead, they can only estimate their sum. (b) Conceptual layout of our multiparameter sensor time-bin multiplexed along a single path. Cascaded measurable parameters $\phi_{1...N}$ are separated by $N-1$ reflectors of reflectivity $R$ in the sensing arm. Both reflected and transmitted light is detected. Output modes are distinguished by their time of arrival at a detector.}
    \label{fig:concept}
\end{figure}

However, if multiple parameters of the same type are along the same path, the sensor will not be able to distinguish between them.
Consider Fig. \ref{fig:concept}(a), in which a sensor has four parallel paths but five parameters to sense. 
The sensor can resolve $\phi_1$ through $\phi_3$, but can only estimate a linear combination of $\phi_4$ and $\phi_5$ instead of $\phi_4$ and $\phi_5$ separately. 
Such a situation may arise in remote sensing, in which an object of interest is physically behind another.
For example, with LIDAR, an oncoming vehicle can be obscured by fog. 
Alternatively, an optical fiber sensor can obtain environmental information (stress, strain, sound, or temperature) as a function of length along a single fiber. 
This information is useful in monitoring oil and gas pipelines and similarly sensitive pieces of infrastructure.
In both cases, there are sensors with only one spatial path, but multiple parameters cascaded along that path. 
Existing multiparameter estimation protocols do not fully explore our cascaded scenario, and both advantage \emph{A} and \emph{B} must be verified in this architecture.

To solve this issue, we borrow the concept of time-bin multiplexing from optical communications \cite{Tucker1988}.
In there, independent signals are sent over the same path as sequential pulses with no overlap. The detector can then demultiplex signals based on their time of arrival.
For a quantum metrological application, consider a state propagating down a single path, and some fraction of that mode is split or reflected after each parameter. The state's time of arrival at the output can help distinguish between each parameter. 
This process bears resemblance to a technique with optical fiber sensors called phase-sensitive optical time domain reflectometry \cite{Redding2020, Kirkendall2004, Lu2019, Wu2019}. There, an optical fiber has $N$ weak Bragg reflectors along its length, dividing the sensor into discrete regions. An environmental variable, such as temperature or strain, changes the local refractive index of the fiber and gives a local phase shift in those regions. After a pulse of light enters the fiber, the reflected light is compared to a reference for a phase measurement, and the time of arrival is mapped to distance along the fiber.
Here we analyze a similar situation in the context of quantum metrology.

In this work, we formulate the theory of cascaded multiparameter estimation within the framework of quantum metrology. In contrast with the multimode sensors \cite{Gorecki2022, Hou2020, Polino2019, Proctor2018, You2017, Zhang2017}, our concept can resolve multiple parameters that are inherently mixed along a single path through time-bin multiplexing. In our sensor, we only have two input and output points, but will estimate $N$ phase shifts.
The enabling idea is that squeezing enhances all information from the sensor, including the correlations between parameters. 
Since these correlations exist for both classical and quantum measurements, the correlations do not inhibit quantum resources from showing a maximal quantum enhancement.
As an example, we model a Mach-Zehnder interferometer with multiple, cascaded phase shifts and a quantum enhancement factor of approximately 7 (or 1.76 given 3 dB of loss).
Moreover, we find that using a coherent state with squeezed vacuum gives a unique advantage in obtaining distinguishing information on the phases while still preserving squeezing in the detected modes. 
These states have an inherent advantage of high power and sensitivity for making a practical sensor, as demonstrated by gravitational wave interferometers \cite{Aasi2013}. 
Further classical modeling suggests how the sensor could scale linearly to 100 or more phase shifts. 
Our approach to multiparameter quantum metrology leverages time-bin multiplexing to resolve parameters along a single path, enabling applications in remote and fiber-based sensing of temperature, strain, acoustics, or seismic activity with greater precision.

\section{Quantum Estimation Formalism}
In this section, we describe the challenges in multiparameter estimation. We explore how current literature overcomes these challenges, and how our cascaded scheme takes a different approach.

Quantum metrology commonly uses the quantum Fisher information, $F$, to quantify the sensitivity of the output state of a sensor. 
For a single parameter, a lower bound on the variance for measuring the parameter $\phi$ is determined by the quantum Cramér-Rao bound \cite{Helstrom1969}:

\begin{equation}
    \label{eq:CramérRao}
    \Delta^2\tilde{\phi} \geq \frac{1}{F}\,,
\end{equation}
where $\tilde{\phi}$ is an unbiased estimator of $\phi$. More generally, with multiple parameters, the quantum Fisher information matrix represents the amount of information obtainable on each parameter with diagonal elements, $F_{ii}$. 
The off-diagonal elements, $F_{ij}$, instead represent the degree of correlation between parameters $i$ and $j$. 
Because $F$ is now a matrix for a vector of parameters $\vec{\phi}$, the Cramér-Rao bound also generalizes to \cite{Ragy2016}

\begin{align} \label{eq:MultiCramérRao}
    & Cov(\tilde{\phi}) \geq \boldsymbol{F}^{-1}, \\
    \label{eq:MultiCramerRao_b} & (\boldsymbol{F}^{-1})_{ii} \geq \frac{1}{F_{ii}}\,,
\end{align}
where $Cov(\tilde{\phi})$ is the covariance matrix for the parameters in $\vec{\phi}$. In other words, the diagonal elements of the inverse of the Fisher information matrix gives a lower bound on the measurement variances of $\tilde{\phi}$ . 
Equation~\eqref{eq:MultiCramérRao} is the direct analog of Eq. \eqref{eq:CramérRao} in the multiparameter framework \cite{Helstrom1969}.
The added subtlety is in Eq.~\eqref{eq:MultiCramerRao_b}, which considers the correlation between parameters \cite{Crowley2014}. 
Taking the inverse of $F$ means that off-diagonal elements in $F$ will  increase the measurement variances of each estimator.

If one wishes to measure cascaded parameters that would otherwise be indistinguishable [e.g. $\phi_4$ and $\phi_5$ in Fig. \ref{fig:concept}(a)], one must introduce a non-commuting operator in between them. The partial reflectors in Fig. \ref{fig:concept}(b) serve this purpose through time-bin multiplexing. Although some correlations between parameters remain, we will show that these correlations do not prohibit a sensor from demonstrating a quantum advantage in sensitivity. 
Because these correlations are classical, an optimal choice of squeezed quadratures could enhance the phase-sensitivity of every mode. This application of squeezing would multiply every term in the Fisher information by the same squeezing factor, resulting in a quantum advantage.
The equal enhancement is true for the examples we have investigated, but even after our analysis it remains an open question whether or not it holds more generally. Overall, despite the measurement being unable to saturate Eq.~\eqref{eq:MultiCramerRao_b}, there is still potential for a quantum enhancement in sensitivity.
Therefore, our goal is to show a quantum advantage in sensitivity by a factor of $e^{2r} \approx 7$ (or 1.76 given 3 dB of loss) for multiparameter estimation, despite the remaining correlations.

In principle, one could diagonalize the Fisher information matrix to remove these correlations with a matrix $\boldsymbol{D}$, but one would then have a measurement basis of $\boldsymbol{D}\cdot \tilde{\phi}$. A mixed measurement basis is undesirable for a cascaded sensor; see Appendix G for a simple example. Moreover, Appendix G shows that our figure of merit for sensitivity, $Tr(F^{-1})$, remains constant under transformation. Therefore, diagonalization offers no overall improvement in sensitivity. 

There are some differences between analyzing the quantum Fisher information and the classical Fisher information. Given an input state and a sesnor layout, one can calculate the quantum Fisher information regardless of how one would detect the output state and estimate the parameter. The classical Fisher information instead analyzes a physical signal from an assumed detection method. If the classical Fisher information matches the quantum Fisher information, then the chosen detection method is optimal. This paper considers the use of a pure Gaussian state with no losses and with large displacement [$|\alpha| \gg sinh(r)$]. With these assumptions, phase sensing with homodyne detection indeed saturates the quantum Fisher information \cite{Ataman2019}. This equality remains true even in our sensor because of the linearity in Fisher information from separate detection events (i.e. separate reflections). Each homodyne detector must still measure the squeezed quadrature of the output state. Therefore, in our model, the classical Fisher information is equal to the quantum Fisher information.

As with any phase estimation procedure, losses will adversely affect the sensor performance. A squeezed measurement with total system efficiency $\eta$ will add noise into the state so that the original quadrature noise $\Delta^2X_0$ becomes \cite{LvovskyPhotonicsch5}
\begin{equation}
    \langle \Delta^2 X \rangle = \eta\langle \Delta^2 X_0 \rangle + (1-\eta),
\end{equation}
where vacuum noise is normalized to 1. In the limit of infinite squeezing, the achievable quantum enhancement becomes $(1-\eta)^{-1}$. An advantage of this state is that losses cannot eliminate squeezing completely. See Appendix B for more details.


\section{Time-Multiplexing Solution}
A simplified schematic of the time-bin multiplexing detection process is illustrated in Fig. \ref{fig:concept}(b). 
A pulsed state is sent down a path with $N$ parameters, each separated by an element of power reflectivity $R$ and transmission $T = 1-R$. 
Detectors resolve the parameters $\vec{\phi}$ using the time of arrival of each reflection back at the input. 
The first reflection of the state arrives first, giving an estimate on $\phi_1$. 
The second reflection gives an estimate on $\phi_2$, and the process continues to $\phi_{N-1}$. 
A detector also sees the final transmitted mode, giving an estimate on $\phi_N$. 
The reflectors do not represent a lossy operation because each reflector is effectively a beamsplitter where both of the output modes are detected. 
Thus a single input mode is split into many different temporal modes along the same path and time-resolved detection of each reflection enables separate estimates on each parameter.

The single pulsed coherent (or displaced) state is the optimal classical probe for this system. Any additional light would only add ambiguity on which reflection interacted with which phase, which increases uncertainty. The ambiguities also appeared for various quantum states with no displacement. Thus the optimal quantum state must also include a displacement, with added squeezing to include quantum resources. Therefore, in determining a quantum enhancement in sensitivity, we compare performance between a displaced (classical) state and a squeezed state with the same displacement.

We outline a statistical approach to time-multiplexed cascaded sensing, including the effect of squeezed (and antisqueezed) quadratures.
The homodyne detectors give a series of orthogonal quadrature measurements $\vec{X}(k | \vec{\phi})$ and $\vec{Y}(k | \vec{\phi})$ for each time bin $k$.
These measurements are used in a series of estimators $\tilde{\phi}(\vec{X}, \vec{Y})$, which give an unbiased estimate on  the parameters $\vec{\phi}$.
Because light is reflected multiple times, an estimator $\tilde{\phi}_i$ may depend on all quadrature measurements $k = 0$ through $n$.
More importantly, the estimator variance $\Delta^2\tilde{\phi}$ is determined by a linear combination of quadrature measurement uncertainties through error propagation:
\begin{equation}
    \label{eq:estimator_variance}
    \Delta^2\tilde{\phi}_i = \sum_{j = 0}^{length(\hat{X})}{b_j \frac{\Delta^2\hat{X}_j}{(\partial \hat{X}_j / \partial \phi_i)^2}+c_j \frac{\Delta^2\hat{Y}_j}{(\partial \hat{Y}_j / \partial \phi_i)^2}}.
\end{equation}
The weights $b_j$ and $c_j$ will depend on the exact sensor configuration and on which quadratures are squeezed. The initial vector of squeezing angles $\vec{\chi}$ can be optimized so that the antisqueezed quadratures have zero weight.
Thus, if the quadratures with nonzero weights are squeezed by a factor of $e^{-2r}$, then each measurement variance $\Delta^2 \tilde{\phi}_i$ will also be reduced by a factor of $e^{-2r}$.
While the procedure is complicated by correlated parameters, the core principle of squeezing-enhanced phase estimation remains unchanged.

\begin{figure*}[t]
    \centering
    \includegraphics[keepaspectratio, width=\textwidth]{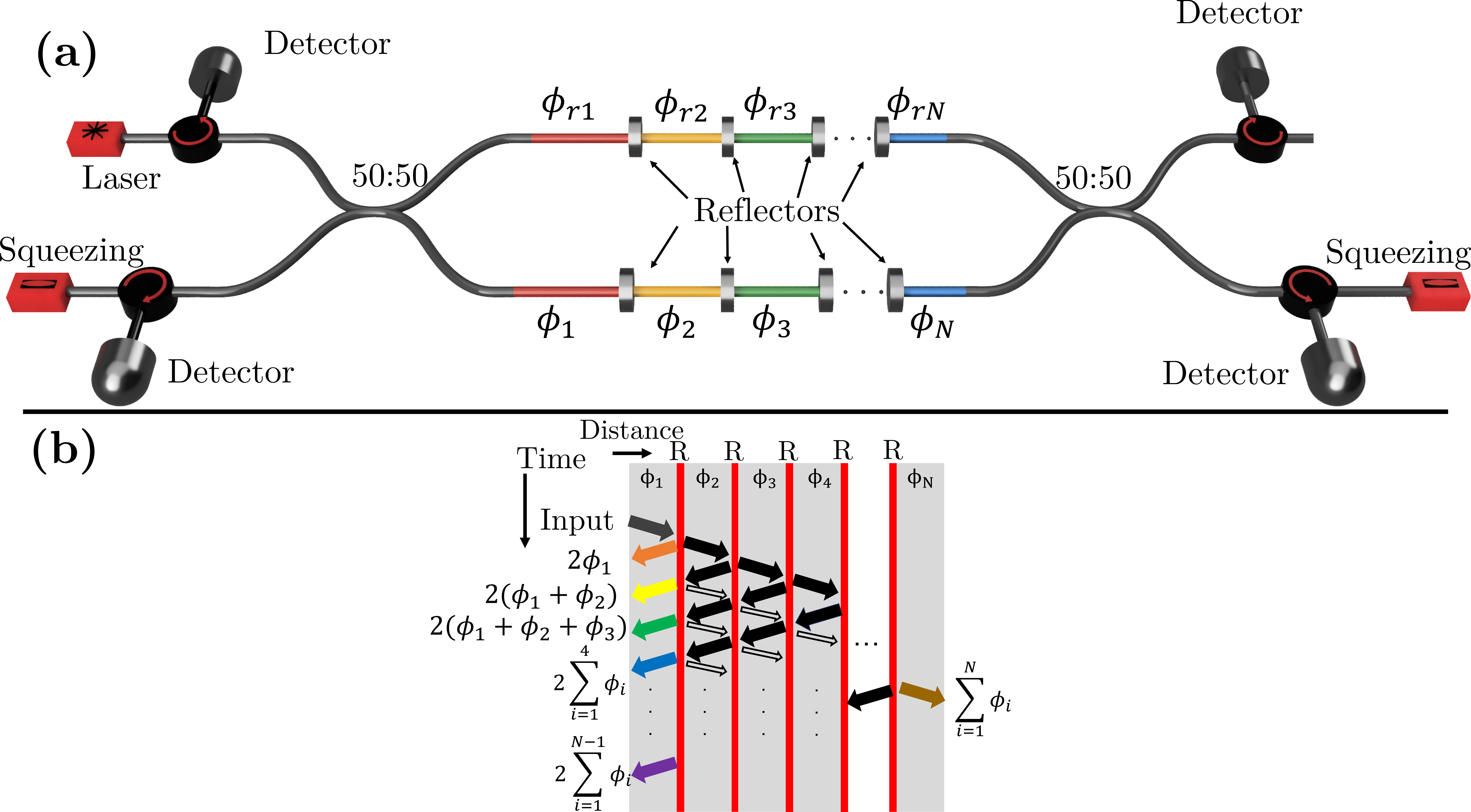}
    \caption{Interferometer for cascaded phase shifts following the time-bin multiplexed concept. With an added reference arm and a 50:50 beamsplitter, the sensor is sensitive to $\phi_i - \phi_{r,i}$. It operates bidirectionally, with input and output on both sides. We choose to input a pulsed coherent state (laser) with pulsed squeezed vacuum in order to show a quantum enhancement in phase sensitivity in a practical setting. Local oscillator modes for each homodyne detector are implied and not shown. (b) Temporal dynamics of the sensing arm of such a sensor, with space on the horizontal axis and time on the vertical. One input (gray) is split into several output modes (colored arrows), each containing information on the listed parameters. The hollow arrows indicate secondary reflections, which can be neglected for $R \ll 1$.}
    \label{fig:layout}
\end{figure*}

For an optical implementation, the parameters are phase shifts within a Mach-Zehnder interferometer with sensing and reference arm, as shown in Fig. \ref{fig:layout}(a). 
A coherent pulse and squeezed vacuum are combined on a beamsplitter, forming a displaced squeezed state inside the interferometer. 
Both sensing and reference arms have evenly spaced reflectors of power transmission $T$ or reflectivity $R = 1-T$, separating the arms into $N$ regions. 
We associate a phase shift $\phi_i$ with each region between reflectors in the sensing arm. 
The reference phases $\phi_{\rm ri}$ are adjusted to maximize phase sensitivity in measuring each phase shift. 
Circulators reroute the reflections to homodyne detectors. 
Our modeling uses a bidirectional sensor shown in Fig. \ref{fig:mappedModes}, where input states, circulators, and detectors are on both sides of the sensor. The extra input offers greater flexibility in optimization.

The core physics of the sensor is illustrated in Fig. \ref{fig:layout}(b), which shows the timing dynamics of how a single input pulse maps to multiple (colored) output modes in the sensing arm of the interferometer. 
Distance is plotted on the horizontal axis, while time is plotted on the vertical.
Each reflected mode arrives at a different time on the detector, and contains information on the listed combination of phases.  In the statistical description, these labels indicate which components of $\vec{X}(k | \vec{\phi})$ and $\vec{Y}(k | \vec{\phi})$ the estimator $\hat{A}_n$ uses to measure $\phi_n$.
The hollow arrows indicate secondary reflections, which can be neglected if $R \ll 1$. 
This is a valid assumption if $N \gg 1$, since we have indications that an optimal design will have $R \propto N^{-1}$.

An intuitive explanation for how the squeezed state in our proposed sensing scheme can enhance sensitivity can be described by the simple  Mach-Zehnder interferometer with squeezed light. The non-Poissonian photon statistics of squeezed light gives reduced vacuum phase noise, which results in a lower measurement variance. The same argument is true for multiparameter sensors, provided the sensor is able to distinguish between those parameters. The proposed cascaded sensor provides distinguishability by measuring multiple temporal modes from the partial reflectors, each giving a squeezed measurement. There are two conditions here for a quantum enhancement. First, the same input state must be able to give a quantum advantage to all parameters simultaneously. Given the degrees of freedom available in the input state, a global optimization can engineer a simultaneous quantum enhancement. Secondly, all output modes must have the same squeezing strength as the input state, which ensures that each measurement on each mode has reduced vacuum noise.

Maintaining the squeezing strength of all output modes is challenging because any squeezing in the input is spread out, or diluted, across all output modes. 
Each reflection operation is functionally a beam splitter with one of its inputs as vacuum. 
The vacuum noise propagates into the output modes and so only a fraction of the original squeezing strength would be observed. 
In contrast, if both input modes are squeezed, the squeezing strength can be preserved.

Thus a countermeasure to the squeezing dilution is to input multiple squeezed pulses so that the other input port on each beam splitter operation is no longer vacuum. With a few starting vacuum modes and an increasing number of squeezed pulses, the output squeezing strength of each mode will asymptotically approach the input squeezing strength. To synchronize with the cascaded sensor's reflections, these pulses can be timed sequentially with a repetition period $\tau$. To consistently match this repetition period, the reflectors must be uniformly spaced within the sensor. The end result is a system where any initial vacuum noise is suppressed along a squeezed quadrature through constructive interference of many squeezed vacuum pulses.

\section{Multiparameter Sensing Model}

We will show that the quantum enhancement in our schematic holds for continuous-variable Gaussian states, beginning by showing quantum-limited sensitivity in a model interferometer with cascaded phase shifts as the parameters.
We compare the optimal sensitivity of a squeezed vacuum + coherent state input to that of a coherent state of the same energy, for both two- and three-phase interferometers. The results of the three-phase interferometer suggest that the sensing procedure can scale to interferometers with many more phases.

Our sensor model uses a matrix formulation of Gaussian states and Fisher information commonly used in continuous-variable quantum information \cite{Adesso2014, Serafini2017}.
The relevant matrix definitions and equations for Fisher information are listed in Appendix A. 
The supporting code is accessible at the GitHub repository \cite{github}.
In our notation, $\hat{\boldsymbol{B}}_{i,j}(T)$ represents a beam-splitter operation between modes $i$ and $j$ with a splitting ratio of $T$ and $\hat{\boldsymbol{P}}_{i}(\phi)$ represents a phase shift operation on mode $i$ with parameter $\phi$.
These matrix definitions are used to propagate a multimode Gaussian state through the sensor with a unitary matrix $\boldsymbol{\hat{U}}$ and to calculate and optimize the Fisher information.
In Appendix A, we also use the notation to  re-prove the quantum advantage in sensitivity in an example (single-phase) Mach-Zehnder interferometer. 

For the current sensor, the relevant figure of merit is the ratio of total phase variances $\sum_{i}{\Delta^2\tilde{\phi}_i } $ with classical light and with quantum light:

\begin{equation}
    \label{eq:Q}
    Q = \frac{\sum_{i}{\Delta^2\tilde{\phi}_{i, \rm classical}}}{\sum_{i}{\Delta^2\tilde{\phi}_{i, \rm squeezed}}},
\end{equation}
where we call Q the quantum advantage in measurement variance. We consider the “classical” case where the squeezing parameter $r = 0$ and the quantum case where $r = 1$. Because we make the approximation that $|\alpha^2 \gg sinh^2(r)$, any added energy from squeezing is negligible in comparison to the coherent state. Thus, the classical and quantum cases have approximately equal energy for a fair comparison in sensitivity. 

Given the use of squeezed light, $Q$ will ideally be equal to the degree of squeezing of the input light, $e^{2r}$, where $r$ is the squeezing parameter.
State-of-the-art squeezing sources can achieve about an order of magnitude of quadrature squeezing, so we choose $r = 1$ for our simulations \cite{Mehmet2010, Vernon2019, Eberle2011}. Given $r = 1$, we expect a maximum sensitivity improvement of $e^{2} = 7.39$.
We also verified in Appendix C that our numerical results scale exponentially with $r$, meaning that any $r > 0$ will result in the expected quantum advantage of $e^{2r}$.

A known issue with the Fisher information is the freedom in the definition of a phase shift. 
One can consider imparting a phase $\phi$ on a single mode, $\hat{P}(\phi)$, or symmetrically imparting $-\phi/2$ and $\phi/2$ onto two modes, $\hat{\boldsymbol{P_{\rm s}}}(\phi)$.
The two representations give different prefactors \cite{Jarzyna2012}, which are not important since they cancel out when comparing the performance of classical and quantum states.  
A phase can only be measured relative to a reference, and there is no such thing as an absolute (single-mode) phase shift in a two-mode interferometer \cite{Jarzyna2012}. 
Because the Mach-Zehnder interferometer has a reference arm, $\hat{\boldsymbol{P_{\rm s}}}(\phi)$ is the correct operator to use there.

While the formalism of Gaussian states and Fisher information remain largely the same for multiparameter estimation, its increased complexity requires some simplification.
A major reduction in complexity comes from using the single-mode phase shift operator.
In the simple, single-phase Mach-Zehnder interferometer, one can obtain the same expressions for Fisher information (with a different prefactor) by simply replacing the Mach-Zehnder unitary matrix with the single-mode phase shift $\hat{\boldsymbol{P}}(\phi)$. 
This situation has a single spatial mode where the input is a displaced squeezed state and represents propagation only through the sensing arm of the interferometer. 
Strictly speaking, these phase shifts are relative to the local oscillator in homodyne detection instead of the reference arm. 

To understand this simplification,  consider the full interferometer with the reference arm, yielding two spatial modes but with single-mode phase shifts. The order of operations between these spatial modes is only important at and after the point of interference. Thus, despite reflections, one can order the phase shifts in separate arms sequentially, and multiply those matrices. The result is a set of diagonal matrices of $e^{i \phi_i} $ and $e^{i \phi_{r,i} }$ for each section $i$ in the arms of the interferometer. Taking out a global phase $exp(\frac{i}{2}(\phi_i-\phi_{r,i}))$ yields the symmetric phase representation. The global phase commutes with all matrices and can be neglected.  
Thus, in the multi-phase interferometer analysis to follow, it is sufficient to consider propagation only in the sensing arm, which halves the required number of simulated modes.

Modeling an interferometer with multiple phases separated by partial reflectors requires careful accounting of the number of spatiotemporal modes $M$ in the system and when these modes interact. 
These modes include both reflection and transmission through partial reflectors, which can form weak cavities. 
While closed-form expressions exist for the phase accumulation and transmission through a cavity \cite{hecht2002}, these are time-averaged and do not capture the timing dynamics needed for our sensor.
For modeling a single-sided sensor with a perfect mirror, the total number of modes will be
\begin{equation}
    M = k(N-2) + 2,
\end{equation}
where $N > 2$ is the number of phases; $k > 0$ is the number of reflections each temporal mode can undergo. 
For a bidirectional sensor with input and output on both sides, 
\begin{equation}
    M = (k+1)(N-2)+1
\end{equation}

In principle $k$ should be infinite for any sensor with more than one partial reflector, since they form a cavity. 
For computational purposes, we restrict $k = 1$ for the classical analysis with many phases, and $k = 7$ for modeling two- or three-phase interferometers with squeezed light.

\begin{figure}[t]
    \centering
    \includegraphics[keepaspectratio, width=0.9\columnwidth]{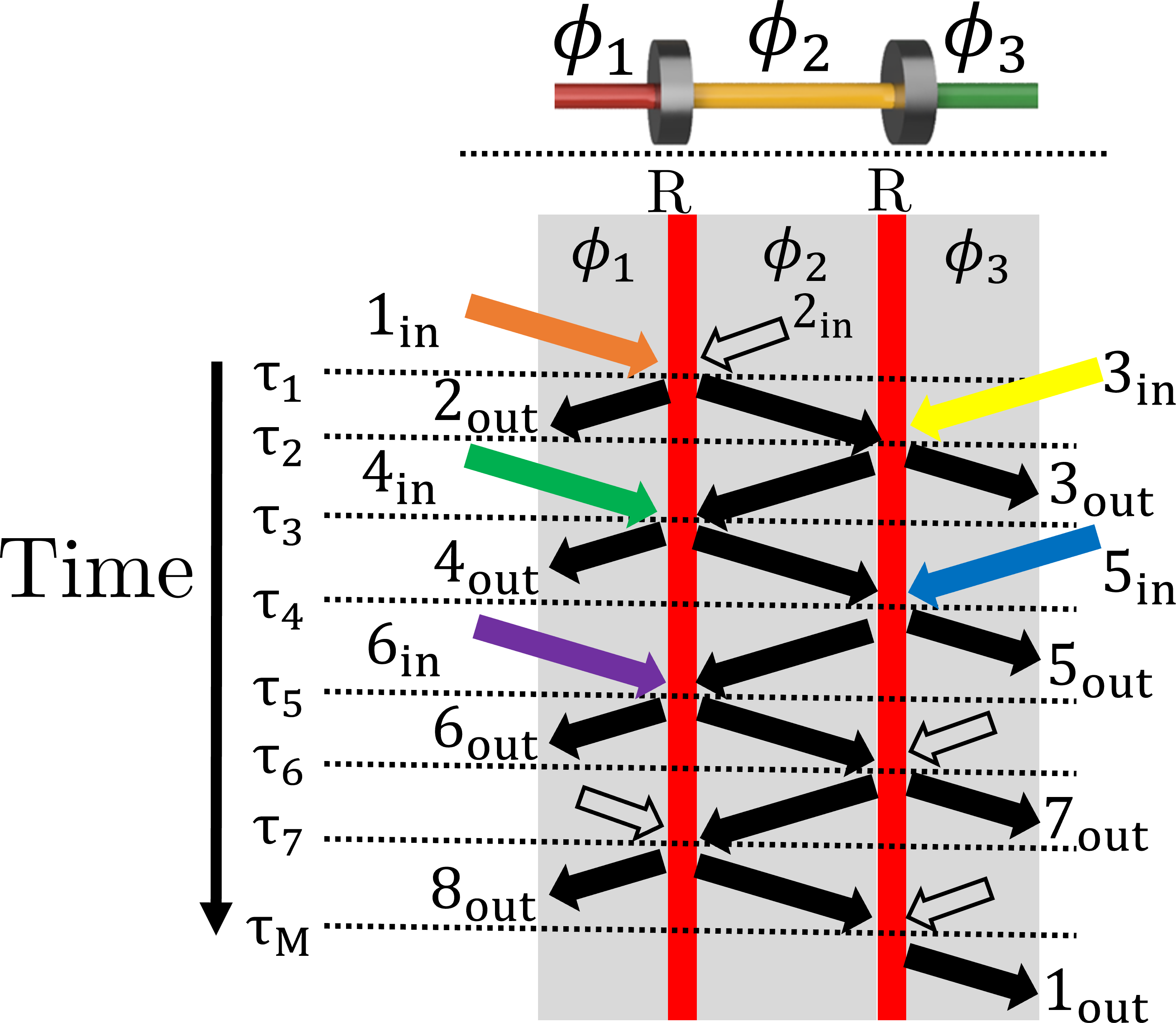}
    \caption{(Top) Schematic of a three-phase sensor with multiple inputs. (Bottom) A visualization of the timing dynamics for probing a three-phase sensor with multiple inputs, as used for the construction of the unitary matrix of the three-phase sensor. In this example, a sensing arm has each temporal mode tracked through each of its phase shift and reflection operations, shown as a function of time. By tracking $k = 6$ reflections from the first input, the input temporal modes (colored arrows) are mapped to eight output modes (black arrows). Five input pulses are considered, with the remaining input modes (hollow arrows) as vacuum.}
    \label{fig:mappedModes}
\end{figure}

A series of beam-splitter (reflector) and phase-shift operations map $M$ input modes onto $M$ output modes in a $2M \times 2M$ unitary matrix. 
Fig. \ref{fig:mappedModes} helps visualize this map for an example with three phases, six reflections, and eight modes in discrete time steps. 
Again, the horizontal axis is distance, while the vertical axis is time.
For step $\tau_1$, we apply $\phi_1$ to mode 1, $\phi_2$ to mode 2, and mix modes 1 and 2 on a beam-splitter. 
For step $\tau_2$, we apply $\phi_1$ to mode 2, $\phi_2$ to mode 1, and $\phi_3$ to mode 3, then mix modes 1 and 3 on a beam-splitter. 
Since the modes are reflected here, the beam-splitter operation is  $\hat{\boldsymbol{B}}_{1,3}(1-T)$ instead of $\hat{\boldsymbol{B}}_{1,3}(T)$. 
Mode 1 continues to reflect inside the interferometer until the end, when the set number of reflections $k$ is reached. 
Thus, we obtain the final unitary as a multiplication of phase shifts and beamsplitter operations for each active mode at each time, where phase shifts act before beamsplitters.

Because we track a finite number of reflections, there is some truncation loss on the final output mode(s) of the sensor. In the present example, mode 1 undergoes a loss transformation instead of the last reflection. 
The truncation operation is equivalent to placing a fictitious beam splitter in that mode's path before detection and tracing over the reflected mode \cite{Ono2010}. See Appendix B for specific implementation, and on how loss and noise generally affect the sensor performance.

Having obtained the output state, the quantum Fisher information matrix for a Gaussian state {${\vec{R}, \boldsymbol{\sigma}}$} with respect to parameters $\phi_i$ and $\phi_j$ is \cite{Safranek2019}
\begin{equation}
    \label{eq:FIMatrix}
    \boldsymbol{F}_{ij} = 2 \frac{\partial \vec{R}^T}{\partial \phi_i}\cdot\boldsymbol{\sigma}^{-1}\cdot\frac{\partial \vec{R}}{\partial \phi_j} + \frac{1}{4} {\rm Tr}[\boldsymbol{\sigma}^{-1}\cdot\frac{\partial \boldsymbol{\sigma}}{\partial \phi_i}\cdot\boldsymbol{\sigma}^{-1}\cdot\frac{\partial \boldsymbol{\sigma}}{\partial \phi_j}].
\end{equation}
From the Fisher information matrix, we can get a lower bound on the sensitivity of measuring N phases with our sensor, which is the main result of this work. 
The sensitivity of each phase is constrained by the multiparameter Cramér-Rao bound with the inverse of the Fisher information matrix in Eq.~\eqref{eq:MultiCramérRao}. 
For a scalar value to optimize, we take $Tr(Cov(\vec{\phi}))$

Given the need to maximize distinguishing information, it is nontrivial to perform a squeezing-enhanced measurement on N parameters simultaneously. 
With no prior phase information, one is equally likely to antisqueeze the necessary  quadrature as one is to squeeze it, which on average would give increased noise. 
A global optimization over all system variables ensures that every mode is squeezed in the correct quadrature such that all parameters are measured with reduced noise, yielding an enhanced sensitivity. 
Interpreting Eq. \eqref{eq:FIMatrix}, maximum phase information requires the derivative of the output state's mean vector to be in the direction of the output state's squeezed quadrature. 

Equation~\eqref{eq:MultiCramérRao} is the total phase variance for a measurement in the sensor and is the figure of merit by which we optimize the sensor's sensitivity. 
With squeezed light, the sensitivity depends on each input pulse's relative phase $\theta_i$, squeezing angle $\chi_{i}$, and the value of each phase in the sensing arm $\phi_{i}$. 
We assume that the coherent state energy $|\alpha_i|^2$ is identical in every instance and much larger than the squeezed vacuum energy. 
The reflector transmission $T$ is also a free variable, but the results are shown as a function of $T$ in order to demonstrate the physics of how transmission affects measurement variance.
With the data of variance versus $T$, one can choose an optimal value of $T$ for a given sensor.
Due to the many degrees of freedom in the sensor, a differential evolution algorithm was used to optimize its sensitivity \cite{Storn1997}. 
Unless otherwise stated, the cost function is $\rm Tr(F^{-1})$. 

The simplest multiparameter case is with two phases and two possible input directions. The unitary matrix for the sensor is

\begin{equation}
    \label{eq:twoPhaseUnitary}
    U = \hat{\boldsymbol{P}}_1(\phi_2) \cdot \hat{\boldsymbol{P}}_2(\phi_1) \cdot \hat{\boldsymbol{B}}_{1,2}(T) \cdot \hat{\boldsymbol{P}}_2(\phi_2) \cdot \hat{\boldsymbol{P}}_1(\phi_1),
\end{equation}
where the subscripts refer to the mode on which the matrix operates. 
Given squeezed vacuum inputs in each mode of strength $r$, we find the quantum Fisher information matrix to be

\begin{equation}
    \label{eq:twoPhaseSqVacFI}
    F = 4 \sinh ^2(2r) \left(
\begin{array}{cc}
 2-T & T \\
 T &  2-T  \\
\end{array}
\right),
\end{equation}
(for $\phi_1=\phi_2=0$). 
The term $4\sinh^2(2r)$ is factored out, leaving only the effect of the reflector transmission inside the matrix. 
The factorable term scales as the square of the average photon number of the input states, $\bar{n} = \sinh^2(r)$, indicating Heisenberg scaling of the phase sensitivity. 
Note that this state has no displaced component, and so homodyne detection will not be optimal in this case.
Moreover, the output covariance matrix under the same optimized conditions is
\begin{equation}
    \label{eq:twoPhaseSqVacCov}
    \boldsymbol{\sigma}_{\rm out} = \left(
\begin{array}{cccc}
 e^{2r} & 0 & 0 & 0 \\
 0 & e^{-2r} & 0 & 0 \\
 0 & 0 & e^{2r} & 0 \\
 0 & 0 & 0 & e^{-2r} \\
\end{array}
\right)\,.
\end{equation}
Now, if mode 1 is instead a displaced coherent state with amplitude $|\alpha \gg 1$ and mode 2 is still squeezed vacuum, we can neglect the second term in Eq.~\eqref{eq:FIMatrix}. In this case, the optimized Fisher information is approximately

\begin{equation}
    \label{twoPhaseFI1}
    F = 4\alpha^2 e^{2r}\left(
\begin{array}{cc}
  4-3T & T \\
 T &  T \\
\end{array}
\right)\,.
\end{equation}
Now the term $\alpha^2 e^{2r}$ factors out, showing that the measurement scales with the coherent state photon number $|\alpha|^2$ and is further enhanced by squeezing with the term $e^{2r}$. 
This is achieved with the same output covariance matrix. 
In summary, both squeezed vacuum and displaced squeezed states show their maximum potential quantum enhancement in sensitivity for the two-phase interferometer, provided that both input modes are squeezed. 

\section{Numerical Results}

For more complicated interferometers, we found no consistent analytic 
method that demonstrates a quantum-enhanced phase sensitivity. Such a solution requires the assumption that an input state exists that maximally squeezes all optimal output quadratures simultaneously. Instead, we use numerical methods to prove the existence of the optimal input state for some - but not all - values of $T$. Our simulations focus on interferometers with $N = 2$ or $N = 3$ phases due to the difficulty in performing a global optimization over many (15+) free variables in a basis with many (12+) spatiotemporal modes. Despite the complexity, one can implement the maximal quantum enhancement in sensitivity of $e^{2r}$.

\begin{figure}[t]
    \centering
    \includegraphics[keepaspectratio, width=0.9\columnwidth]{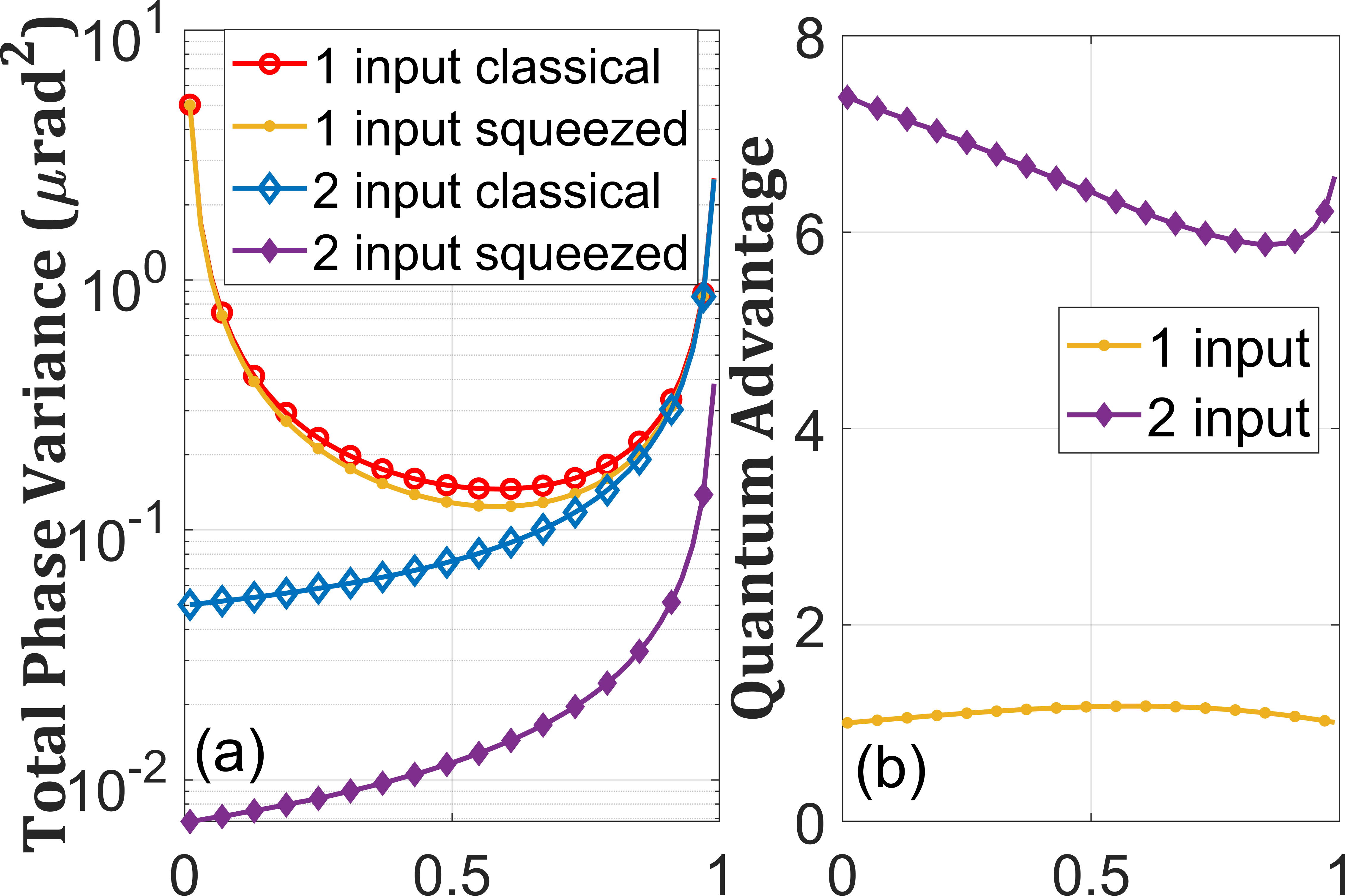}
    \caption{Modeling results of measurements in a two-phase bidirectional interferometer with coherent + squeezed vacuum states. Input parameters were set to $\alpha = 10^5$ and $r = 1$. Bidirectional input instead used $\alpha = 2^{-1/2} \times 10^5$ to keep total energy constant. (a) Total phase variance of the measurements is plotted as a function of reflector transmission $T$ for one or bidirectional input and for classical or squeezed light. (b) Quantum advantage $Q$ is plotted from the data in (a), showing a substantial advantage from using bidrectional input.}
    \label{fig:twophaseCW}
\end{figure}

We start by modeling a two-phase, bidirectional interferometer with displaced squeezed states. 
We compare the total measurement variance as a function of reflector transmission for four different iterations: one input (passing through $\phi_1$ first) or two inputs (from both directions), and with squeezing turned on and off to represent the quantum and classical case, respectively. 
Results are shown in Fig. \ref{fig:twophaseCW}. 
First, compare the two cases with one input. 
With both classical and quantum input, the total measurement variance diverges at $T = 0$ and at $T = 1$. 
Near $T = 0$, the divergence is due to very little light transmitting through the reflector. Since the signal for $\phi_2$ decreases and noise (from the electromagnetic vacuum) is constant, the signal-to-noise ratio decreases.
Near $T = 1$, distinguishing information on each phase diminishes to zero due to an ambiguity in which either $\phi_1$ or $\phi_2$ caused a phase shift. 
Moreover, in Fig. \ref{fig:twophaseCW}(b), the quantum advantage from a single input is at most 17\% near $T = 0.63$. 
The reflector introduces vacuum noise into the state, effectively reducing the squeezing strength for the measurement.   

The two-input or bidirectional input case shows substantially more promise. 
While the measurement variances still diverge near $T = 1$ for the same reason as before, they do not near $T = 0$. 
At that point, the system is essentially two separate interferometers, each independently measuring $\phi_1$ or $\phi_2$. 
The quantum advantage correspondingly starts at its limit near $T = 0$. 
Interestingly, the quantum advantage remains high for all $T$ because squeezed light is interfering with additional squeezed light at the reflector, instead of vacuum noise. 
Thus there is no effective loss of squeezing.
The quantum advantage for intermediate values of $T$ is slightly lower because the optimal points for measuring $\phi_1$ and $\phi_2$ are different. 
Thus the chosen measurement makes some compromise between the two.
See Appendix F for more details.
Despite the two phases being nearly inseparable in the output state when $T \approx 1$, one can still measure both simultaneously with a substantial quantum advantage. 
The general key here is to avoid any interaction with vacuum noise by introducing additional squeezed light at every reflector such that squeezed light from two directions interfere.

A three-phase bidirectional interferometer is more representative of the general $N$-phase system due to the use of multiple synchronized pulses and multiple reflections. 
To handle these multiple reflections, we tracked at least $k = 7$ reflections for each temporal mode. 
We consider a ``downsampled" case where pulsed squeezed vacuum is input in every mode and some subset of the time a displaced squeezed state enters the interferometer instead. The time between displaced squeezed states is long enough that each fully exits the interferometer before the next one enters, avoiding any ambiguity on when the state entered. 
The squeezed vacuum keeps the interferometer modes squeezed, while the displaced input gives the detectors distinguishing information between the phase shifts. 

\begin{figure}
    \centering
    \includegraphics[keepaspectratio, width=0.9\columnwidth]{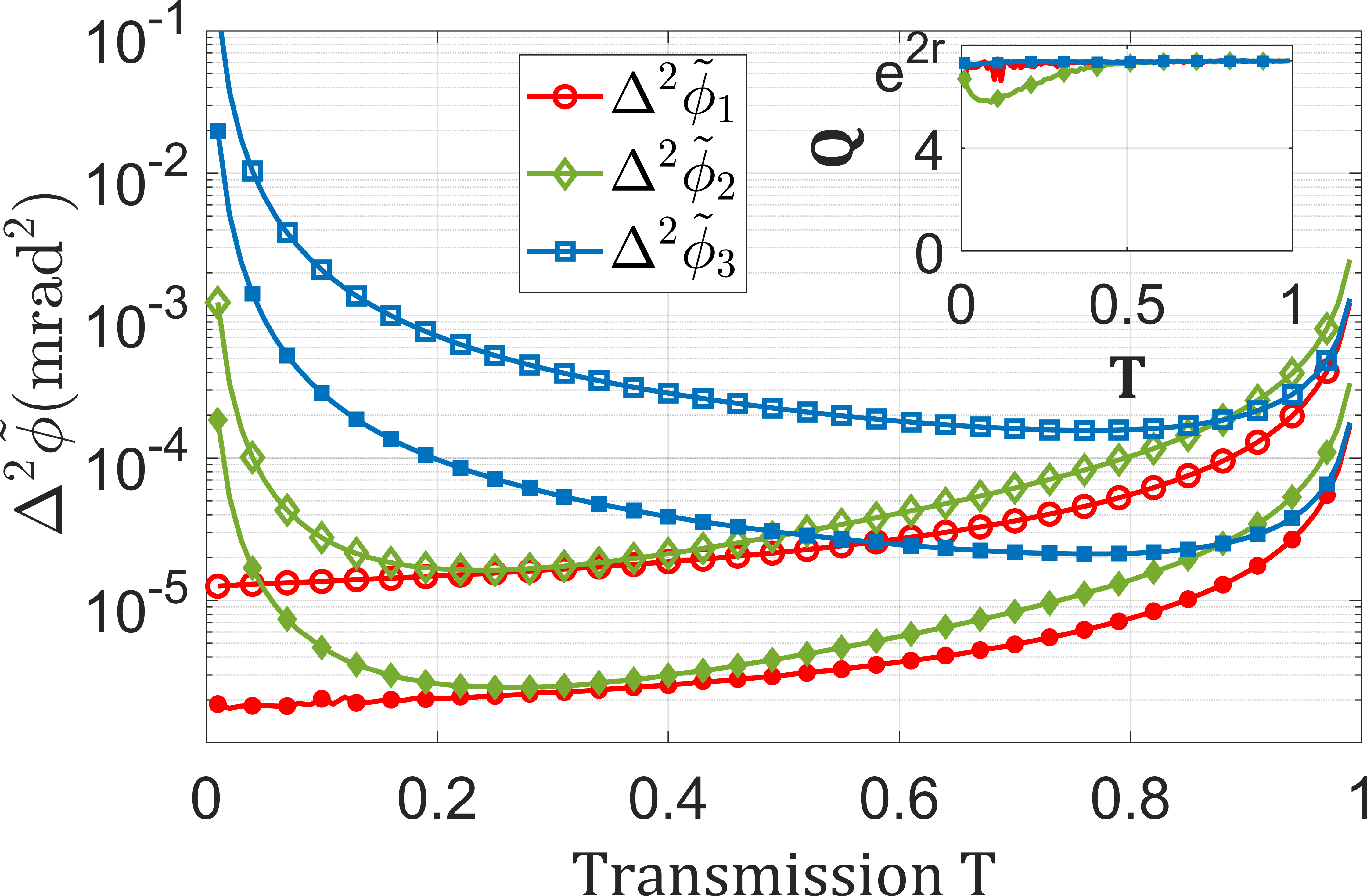}
    \caption{Numerical results for a three-phase interferometer, where a single displaced squeezed state ($|\alpha| = 10^5$) is input into an interferometer with all other modes as squeezed vacuum. The phase variance is plotted for each phase as a function of reflector transmission $T$. The lines with hollow markers represent the performance without squeezed vacuum and only a coherent state, while the lines with solid markers represent the performance with squeezed vacuum and a squeezed coherent state input. Inset is a plot of each phase's quantum advantage $Q$, defined as the ratio of classical variance and quantum-enhanced variance. Further scaling with both classical and quantum photon number can be found in Appendix C.}.
    \label{fig:3phDownsampled}
\end{figure}

Results of the ``downsampled" case are shown in Fig. \ref{fig:3phDownsampled} for a displacement amplitude $|\alpha| = 10^5$ and a squeezing parameter $r = 1$. 
It shows the phase variance for each phase as a function of $T$ both with and without squeezing.
Summing all variances, we find that the lowest total phase variance occurs at $T = 0.62$.
Taking the ratio of these cases for each phase with and without squeezing, the net quantum advantage is within 5\% of the maximal value of $e^{2}$ for $T < 0.62$ and exactly the maximal value for $T \geq 0.62$.  As was the case in the two-phase sensor, a maximal quantum advantage is achievable for the optimal value of $T$.
These results confirm that a full quantum advantage is possible in a time-multiplexed cascaded phase sensor. 
While the time multiplexing tends to dilute any single squeezed state across many output modes, many squeezed states can maintain the maximum level of squeezing inside the interferometer. 
Moreover, to maximize distinguishing information, the downsampled displaced squeezed state provides the needed timing information for the detectors. 
The additional degree of freedom provided by the displaced squeezed state probe drives this dynamic and gives a unique advantage to using this state for such time-multiplexed sensors.

We also analyzed a different case where only displaced squeezed states are used, and the states are input sequentially as shown in Fig. \ref{fig:mappedModes}. See Appendixes D and E for details, and for how such a multiparameter scheme compares with measuring phases individually.

\section{Extension to Many Phases}

Here we extend our  analysis to the case where a sensor has $N$ phases, with some simplifications. The results enable us to discuss the viability of a quantum enhancement with many phases, as well as discuss the multiparameter enhancement - advantage \emph{A} - in the context of a cascaded sensor.

To analyze a sensor with many phases, we make two key simplifications to avoid the computational complexity of a $>100$-parameter global optimization.
First, we use only a coherent state so that the covariance matrix of the state will always be the identity matrix. 
Consequently, the first term in Eq. \eqref{eq:FIMatrix} will be phase independent and second term will always be zero, and so the calculation will not require any optimization over each phase. 
Second, we take $k = 1$ so that we track only the first reflections from each reflector. 
Any reflected light that hits another reflector instead transmits with loss $1-T$.

Let us walk through the simple case of $N = 2$, detecting both reflected and transmitted modes. 
The unitary matrix that describes this sensor is
\begin{equation} \label{eq:OTDR2phaseU}
    U = \hat{\boldsymbol{P}}_1(\phi_2) \cdot \hat{\boldsymbol{P}}_2(\phi_1) \cdot \hat{\boldsymbol{B}}_{1,2}(T) \cdot \hat{\boldsymbol{P}}_1(\phi_1)\,.
\end{equation}
Given an input mean vector $\vec{R}_{\rm in} = (\alpha, 0, 0, 0)$, the output vector is
\begin{equation} \label{OTDR2phaseoutput}
    \vec{R}_{\rm out} = \sqrt{2} \alpha
    \begin{pmatrix}
         \sqrt{T} \cos(\phi_1 + \phi_2) \\ \sqrt{T} \sin(\phi_1 + \phi_2) \\ -\sqrt{1-T} \cos(2\phi_1) \\ -\sqrt{1-T} \sin(2\phi_1)
    \end{pmatrix} \\ \,.
\end{equation}
The resulting Fisher information matrix is
\begin{equation} \label{OTDR2phaseFI}
    \boldsymbol{F}(T, \alpha) = 4\alpha^2
    \begin{pmatrix}
        (4-3T) & T \\
        T & T
    \end{pmatrix} \,.
\end{equation}
Optimal phase sensitivity is obtained when the trace of the inverse of $\boldsymbol{F}$ is minimized. 
This occurs at $T = 0.586$ and gives a total phase variance of $0.73/\alpha^2$.

For more complex sensors, there is actually no need to compute the unitary matrix and output state because the Fisher information matrix has the following pattern:

\begin{align} \label{OTDR_FIpattern}
    &F_{n,n} = 4T^{n-1} \alpha^2, \\
    &F_{i,i} = F_{i+1, i+1} +\alpha^2(16T^{2i-2} - 16T^{2i-1}), \quad i = 1:n,\\
    &F_{i,j} = F_{j,i} = F_{i,i}, \quad j = 1:i-1,\quad i = 1:n\,.
\end{align}
Similar patterns exist for other sensor configurations. For instance, if the layout in Fig. \ref{fig:layout}(a) ended in a perfect mirror instead of a second beamsplitter, $F_{n,n}$ changes to $16T^{2n-2} \alpha^2$. This simple and elegant result shows how similar sensors can be analyzed with the same methods.

We repeated this procedure again in a different scenario where the sensor operates bidirectionally. 
That is, a measurement from one side of the sensor is followed by another pulse from the opposite side. 
The Fisher information changes by averaging itself with a copy that has both columns and rows reversed:

\begin{equation}
    \label{FIotdr_bidirectional}
    F_{\rm bidirectional}(T,\alpha) = \frac{\boldsymbol{F}(T, \alpha) + \boldsymbol{J}\cdot \boldsymbol{F}(T, \alpha) \cdot \boldsymbol{J}}{2}.
\end{equation}
The exchange matrix $\boldsymbol{J}$ is the identity matrix with its rows reversed. 
Functionally, the reversed Fisher information matrix swaps definitions between $\phi_i$ and $\phi_{n-i+1}$, as if the input state started on the other side of the sensor. 

\begin{figure*}[t]
    \centering
    \includegraphics[keepaspectratio, width=\textwidth]{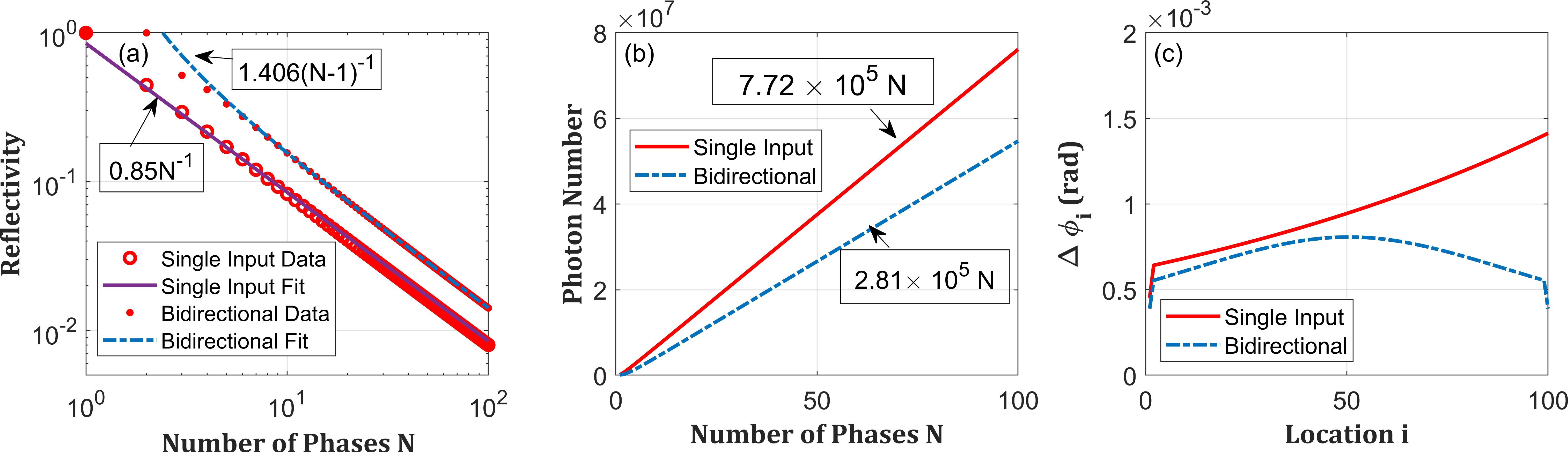}
    \caption{Results of modeling a sensor with up to 100 phases and classical light with two scenarios. The single input uses a single pulse from one side for measurement, while the bidirectional case uses two pulses, one from each side. (a) Plot of the optimal reflectivity $R_{\rm opt} = 1-T_{\rm opt}$ of the partial reflectors as a function of the number of phases $N$ in that sensor. (b) The required power, or photon number, required to reach an average phase variance of $1 mrad^2$, as a function of the number of phases in the sensor. (c) Distribution of the standard deviation of a phase measurement for each phase in the sensor ($N = 100$)}
    \label{fig:OTDRfig}
\end{figure*}

With these patterns, Fisher information matrices were constructed for up to $N = 100$ phases. 
In order to study how the sensor scales with more measurable phases, we found the optimal reflector transmission  $T_{\rm opt}$ for each $N$ that minimizes the total phase variance.
Fig. \ref{fig:OTDRfig}(a) plots $T_{\rm opt}(N)$ for both single and bidirectional cases. 
In each case, the optimal transmission shows clear $1/N$ dependence. 
Consequently, the normalized transmitted power $T^N$ asymptotically approaches 0.45 and 0.24 for the single and bidirectional cases, respectively. 
We then used $T_{\rm opt}(N)$ to determine what power (photon number) in the sensor is required to reach an average phase variance of 1 $mrad^2$. 
Fig. \ref{fig:OTDRfig}(b) shows that this power requirement remains linear in $N$ for both cases, with slopes of $7.72 \times 10^5$ and $2.81 \times 10^5$ photons per added phase for each case. 
Fig. \ref{fig:OTDRfig}(c) also shows the distribution of phase variances for the 100-phase sensor. 
In both cases, the measurement variance is highest where the sensor has the lowest circulating power. 
For the single-sided case, this point is at its opposite end, while it is in the middle for the bidirectional case. 
Overall, the linear scaling of this sensor to 100 phases shows that there are no fundamental limits to this architecture extending to many phases.

In this analysis, some light was lost by the assumption that $k = 1$ so that we track only the first reflections from each reflector. Given reflector transmission $T$ and total number of phases $N$, this loss $L(T, N)$ is
\begin{equation} \label{eq:OTDRloss}
    L(T, N) = 1-T^{N-1}-\sum_{k = 0}^{N-2}T^{2k}(1-T)\,.
\end{equation}

Assuming $T(N) = x/N$ for some proportionality constant $x$, we can take an infinite limit:
\begin{equation} \label{eq:OTDRloss2}
    \lim_{N\to\infty} L(T,N) = \frac{1}{2}e^{-2x}(e^x-1)^2.
\end{equation}
Given the constants of proportionality fitted in Fig. \ref{fig:OTDRfig}(a) for optimal reflectivity, we find that the loss is 16.4\% and 28.5\% for the single-input and bidirectional cases, respectively. While these losses are rather high, including more reflections would add higher-order terms to $T(N)$ and show increased sensitivity per measurement. However, the measurement interval will take $k$ times longer for measuring the $k^{\rm th}$ order reflection, with diminishing returns on sensitivity. Measuring all reflections will be optimal for extracting the maximum amount of information per photon. 

However, if one instead wants to obtain the most information per unit time, the optimal strategy is to only track reflections up to order $k$ before sending the next pulse. This method will be significantly faster than waiting for all detectable reflections, depending on the error tolerance in the end application.

The remaining reflections will interfere with the next measurement, giving small errors that can be mitigated. If $k = 1$, then the total relative error will be proportional  to Eq. \ref{eq:OTDRloss}.  Tracking more reflections with a longer measurement will lower this error to be proportional to $L(T,N)^k$. At sufficiently high $k$, these errors will be lower than shot noise. Additionally, the estimation algorithm can be tuned to accommodate leftover reflections because they are still correlated with the measurable phases. Lastly, one could slightly delay the next probe pulse so that it no longer overlaps with the remaining reflections, and ignore them because they are now outside the detectors’ expected timing window.

\section{Discussion}

With these results, we find that a multiparameter enhancement - advantage \emph{A} - is clearly present, but also paradoxical because these parameters are not individually addressable. Suppose one wanted to estimate a single parameter, $\phi_i$, given all other phases in $\vec{\phi}$ are known. Given the sensor layout, one would have to probe $\phi_i$ in a similar way to the multiparameter measurement. For instance, light must pass through $\phi_{1...i-1}$ to reach $\phi_i$. One may estimate $\phi_i$ marginally better in a single-parameter measurement, but one could estimate all of $\vec{\phi}$ with the same probe and detectors simply by processing data differently. Comparing $N$ single-parameter measurements to one multiparameter measurement of comparable average sensitivity, advantage \emph{A} should be linear in $N$. The disclaimer is that the single-parameter estimation used in the comparison is unrealistic, and the concept of a multiparameter advantage was meant for other situations where parameters are individually addressable.

Advantage \emph{B}, the quantum enhancement, is also present in the system. The expected squeezing factor of $e^{2r}$ appears in both our analytic results for the two-phase interferometer, and in the three-phase interferometer for $T \geq 0.62$.
A likely explanation is that the Fisher information matrix for an optimized cascaded measurement will factor out into two terms:
\begin{equation}
    \label{eq:factorableFI}
    \boldsymbol{F}(T, \alpha, r) = f(\alpha, r)\boldsymbol{G}(T).
\end{equation}
Here, $f(\alpha, r)$ is a global factor representing the power and squeezing factor of the input state. 
The remaining matrix $\boldsymbol{G}(T)$ contains the remaining information on the geometry of the sensor and the measurement. Equation ~\eqref{eq:factorableFI} assumes that the input modes are static, so that the structure of $\vec{R}_{\rm in}$ is unchanged. The values of $\alpha$ and $r$ may change, so long as $|\alpha|^2 \gg sinh^2(r)$.
Given this factorability, the measurement variance for all parameters will scale as $f(\alpha, r)^{-1}$.
Since $f(\alpha, r)$ scales as $|\alpha|^2$ for a classical coherent state $|\alpha \rangle$, and $|\alpha|^2 e^{2r}$ for a coherent + squeezed vacuum state, the factor of $e^{2r}$ is again the quantum advantage in sensitivity. Moreover, the factorability shows that our results are independent of the specific value of $|\alpha|$ and $r$. See Appendix C for further tests with varying $|\alpha|$ and $r$.

Our results demonstrate the potential for a quantum advantage in a cascaded system, but relied on some optimistic assumptions. First and foremost is not having optical loss. While Appendix B illustrates how loss should effectively reduce the squeezing factor, there is another consequence. Depending on the platform of choice, losses will constrain the number of phases one can probe. The advantage of optical fiber is the ultralow propagation loss (0.2 dB / km). Moreover, point-inscribed Bragg reflectors can serve as the partial reflectors and have extremely low loss \cite{Williams2012}. In an extreme case, 100 reflectors introduced 0.01 dB of additional loss \cite{Redding2020}. Other platforms such as free-space optics will have higher losses (~1\%), restricting them to probing a few dozen phases for the same system efficiency. Therefore, this method is best suited for optical fiber sensors.

Another consequence of these ideal assumptions is the rate at which a measurement would saturate the quantum Cramer-Rao bound. A lossless pure state is known to saturate its Holevo Cramer-Rao bound in a single shot \cite{Demkowicz-Dobrzanski2020}. With no fundamental incompatibility between our parameters (only classical correlations), we infer that a single-shot measurement is also sufficient to achieve the quantum Cramer-Rao bound that we report. More realistically, mixed states will take multiple copies to saturate these bounds. This fact creates some tradeoff between measurement bandwidth and precision not seen in classical sensors.

\subsection{Practical Considerations}

Synchronizing many squeezed pulses for constructive interference will be challenging, but feasible. For this purpose, the ratio of timing jitter to pulse duration determines how well successive pulses will overlap. Similar to optical loss, any pulse mismatch will reduce the observed squeezing strength at the output. Therefore, pulse mismatch only needs to be a negligible effect compared to the system’s total loss. For example, given a very optimistic 20\% loss, pulse mismatch should be below 2\%. Borrowing from classical fiber sensors, one can avoid dispersive effects and significantly relax timing requirements by using ns-long pulses \cite{Redding2020, Lu2019}. If using a mode-locked fiber laser, filtered down into a ns pulse, one inherits the exquisite timing jitter of these systems, often less than 1 ps rms \cite{Kim2016}, which gives < 0.1\% mismatch. Alternatively, one can carve out a pulse from amplitude modulation in an EOM or AOM. In this case, timing jitter is mostly determined by the radio-frequency driving electronics, typically about 10 ps  \cite{RFJitter}, giving 1\% mismatch. Either method is sufficient for a first demonstration.

Further modulation is necessary to ensure optimal control of these pulses. Phase drifts inside the interferometer would alter the interference conditions but can be compensated by phase-modulating the squeezing pump and local oscillator. This way, each pulse can have freely varying coherent state phase and squeezing angle. Choosing these angles can be determined from prior interferometer measurements.

Implementing the multiphase MZI will have three key challenges: obtaining a pulsed squeezed light source, maintaining uniform reflector separation, and implementing sufficiently fast feedback to the reference arm. 
For the squeezed source, a suitable pulsed source has been implemented in \cite{Madsen2022} for Gaussian boson sampling. 
To ensure high interference visibility, spacing between reflectors should be uniform within 1\% of the pulse length. This constraint is significantly relaxed by using long, nanosecond pulses, giving a pulse length of around 1 m and a spacing tolerance for the reflectors of about 1 cm.
The feedback circuit to the reference arm must also be fast enough to track drifts in the signals, which typically occur at sub-kHz frequencies. 
Feedback into the reference arm can be as fast as several kHz \cite{Lundin2011, Xie2009, Fritsch1981}, which is sufficient. 
The authors in \cite{Madsen2022} also implemented multiple, independently stabilized fiber loops, which is a functionally similar task to phase estimation. 
Therefore, such feedback can also be implemented in parallel to account for our multiple parameters, and a full experiment is feasible.

We predict that our sensor can be implemented as a quasi-distributed temperature sensor in optical fiber with unparalleled sensitivity. 
Given the thermo-optic coefficient of silica of $4 \times 10^{-6} K^{-1}$, a measurement variance of $1$ mrad, and sensing regions $10$ m long, the shot-noise limited temperature sensitivity will be approximately 10 $\rm \upmu K$. 
The quasi-distributed nature again comes from the multiplexing reflectors, which divide the sensor into discrete regions, and each region can be monitored independently.
While microkelvin-sensitive sensors already exist, they typically have sub-Hz measurement bandwidths \cite{Tan2017} and measure at only one location. In contrast, the fiber-based sensor can have kHz measurement bandwidth, comparable sensitivity, and the capability to measure many regions simultaneously. Its distributed nature allows it to infer temperature gradients, which has its own set of applications in locating hot spots or thermal leakages. Such a sensor could provide additional stability to highly sensitive space-based experiments such as the LISA gravitational wave observatory \cite{Gibert2015}.
The sensor could also monitor other transient material properties, such as strain or vibration, that would also induce a change in optical path length in a material. 
Cascaded phase sensing lays a foundation for many remote sensors with its technique of precision interferometry.

\section{Conclusion}
We have developed the framework for a branch of quantum metrology we call cascaded phase sensing. 
Through time-bin multiplexing, a single sensor can resolve multiple parameters along its length.
Regardless of the correlations between parameters in the Fisher information, squeezed states can be used to achieve a quantum advantage of $e^{2r}$, where $r$ is the squeezing parameter.
Therefore, statistical correlations between parameters \emph{due to a chosen sensor layout} do not prohibit a multiparameter sensor from achieving a quantum enhancement.

We present an example structure featuring evenly spaced reflectors on the sensing path for defining phase sensing regions, and time-bin multiplexing for effective use of multiple pulses of squeezed light. 
Despite the cascaded phases sharing an optical path, we show that squeezed light can give a full quantum enhancement in sensitivity of $e^{2r}$ in phase sensing. 
With a state-of-the-art squeezed source, this enhancement can be about a factor of 7 (or 1.76 given 3 dB of loss). 
While we show this enhancement for two and three phases, we anticipate that the same physics applies to sensors with many more phases, limited by propagation loss. 
The classical analysis of an $N$-phase sensor demonstrates a simple and elegant way to optimize the transmission value of reflectors in these sensors. Through further development, optical time-domain reflectometry and similar remote sensing technologies can directly benefit from the optimization procedure.

Our sensing protocol offers a rich branch of quantum-enhanced sensing to explore, combining multiparameter quantum metrology with remote sensing. 
This methodology applies to any quasi-distributed sensing of parameters that can be mapped to an optical phase shift. 
Applications include calibration of networks in quantum communications, monitoring of structural integrity, geophysical surveying, and in free-space remote sensing.

\begin{acknowledgements}
    We acknowledge useful technical discussions with J. Combes, K. Shalm, M. Mazurek, M. Messerly, S. Libby, R. Mellors, M. Zohrabi, T. Shanavas, D. Wilkerson, S. Diddams, M. Nicotra, and I. Novikova. 
    We gratefully acknowledge funding from ARPA-E, Grant No. DE-AR0001152, NSF CCF Grant No. 1838435, Q-SEnSE: Quantum Systems through Entangled Science and Engineering (NSF QLCI Award OSI-2016244), and the University of Colorado Boulder Cubit initiative.
\end{acknowledgements}

\appendix

\section{Matrix Definitions}\label{appendix:A}
Here we define our Gaussian state basis, the matrix definition of operators, and the calculation for Fisher information.
A thorough treatment for continuous-variable quantum information can be found in \cite{Adesso2014, Serafini2017}.
As a simple example, we benchmark this formulation with the classic example of a (single-phase) Mach-Zehnder interferometer.
A Gaussian state can be represented by a mean vector $\vec{R} = \{x, y\}$ and a covariance matrix $\boldsymbol{\sigma}$. 
The vector $\vec{R}$ represents the amplitude of the state in terms of the quadrature operators $\hat{X}$ and $\hat{Y}$, which are analogous to $\hat{x}$ and $\hat{p}$ of the quantum harmonic oscillator, and are accessible by homodyne detection. 
The matrix $\boldsymbol{\sigma}$ represents the quantum noise of that state, normalized to 1.
For example, a coherent state of amplitude $\alpha$ and phase $\theta$ will have:

\begin{align}\label{eq:coherentstate}
    & \vec{R} = (|\sqrt{2}\alpha|\cos(\theta), |\sqrt{2}\alpha|\sin(\theta))\,, \\
 &\nonumber \boldsymbol{\sigma}_{\rm c} = \boldsymbol{I}_2\,,
\end{align}
where $\boldsymbol{I_2}$ is the $2\times 2$ identity matrix. 
The noise matrix is in natural units, $\hbar = c = 1$, so that vacuum noise has an amplitude of $1$ \cite{Adesso2014}. 
In contrast, a squeezed coherent state may have the same amplitude but a different noise matrix:

    \begin{align}\label{eq:Squeezer}
    & \boldsymbol{\sigma}_{\rm sq} = \boldsymbol{\hat{S}}(r, \chi)\hat{\boldsymbol{S}}^{\dagger}(r, \chi)\,, \\
    & \nonumber \boldsymbol{\hat{S}}(r, \chi) = \\
    & \nonumber \begin{pmatrix}
        \cosh(r)+ \cos(\chi)\sinh(r) & \sin(\chi)\sinh(r)\\
         \sin(\chi)\sinh(r) & \cosh(r)-\cos(\chi)\sinh(r)
    \end{pmatrix} \,.
    \end{align}
Here, $\hat{\boldsymbol{S}}(r, \chi)$ is the single-mode squeezing operator with squeezing strength $r$ and squeezing angle $\chi$. 

These states can also be multimode, representing every spatial or temporal mode in a system. 
In this case, the mean vector and covariance matrix will be of size $2M$ for $M$ modes. 
All single-mode expressions can grow to the appropriate matrix size by direct sum with the identity. 
For example, an $M$-mode covariance matrix in which the first mode is squeezed and the rest are vacuum is expressed as:

\begin{align}\label{eq:eq3}
    \boldsymbol{\sigma}_{\rm multimode} = \boldsymbol{\hat{S_1}}(r, \chi)\hat{\boldsymbol{S_1}}^{\dagger}(r, \chi) \oplus \boldsymbol{I_{2M-2}}.
    \end{align}

Gaussian states can propagate through any linear optics, represented with a unitary matrix. 
To demonstrate how this formalism can show a quantum enhancement in sensitivity, we go through the example of a Mach-Zehnder interferometer (MZI). 
Any interferometers modeled here are a combination of beamsplitters and phase shifters. 
When the beamsplitter and phase-shifter matrices are multiplied together, the resulting unitary represents propagation through the entire sensor. 
The two-mode beamsplitter operator with transmission $T = 1 - R$ we use is $\hat{\boldsymbol{B}}(T)$:

\begin{equation} \label{eq:BS}
    \hat{\boldsymbol{B}}(T) = \left(
    \begin{array}{ccccc}
     \sqrt{T} & 0  & \sqrt{1-T} & 0 \\
     0 & \sqrt{T}  & 0 & \sqrt{1-T} \\
     -\sqrt{1-T} & 0  & \sqrt{T} & 0 \\
     0 & -\sqrt{1-T}  & 0 & \sqrt{T} \\
    \end{array}
    \right).
\end{equation}
The single-mode phase-shift operator for a phase $\phi$ acting on a single mode is $\hat{\boldsymbol{P}}(\phi)$\cite{Serafini2017}:

\begin{equation} \label{eq:PS1}
    \hat{\boldsymbol{P}}(\phi) = \left(
    \begin{array}{cc}
    \cos (\phi ) & -\sin (\phi ) \\
    \sin (\phi ) & \cos (\phi ) \\
    \end{array}
\right).
\end{equation}
One may also consider a symmetric, two-mode phase-shift operator, $\hat{\boldsymbol{P_{\rm s}}}(\phi)$ operating on the sensing arm and the reference arm: 

\begin{equation} \label{eq:PS2}
    \hat{\boldsymbol{P}_{\rm s}}(\phi) = \left(
    \begin{array}{ccccc}
     \cos (\phi/2) & -\sin (\phi/2)  & 0 & 0 \\
     \sin (\phi/2) & \cos (\phi/2)  & 0 & 0 \\
     0 & 0  & \cos (\phi/2) & \sin (\phi/2) \\
     0 & 0  & -\sin (\phi/2) & \cos (\phi/2) \\
    \end{array}
    \right)
\end{equation}
Given that a Mach-Zehnder interferometer is sensitive to the phase difference between its two arms, these two phase-shift operations are physically equivalent. 
Thus the Mach-Zehnder interferometer can be expressed as the unitary:
\begin{equation}
    \label{eq:UMZI}
    \boldsymbol{U}_{\rm MZI} = \hat{\boldsymbol{B}}(1/2) \cdot \hat{\boldsymbol{P}_{\rm s}}(\phi) \cdot \hat{\boldsymbol{B}}(1/2).
\end{equation}
For a state $\{R, \sigma\}$, the output state is, in general:

\begin{align} \label{eq:propagation}
   & \vec{R}_{\rm out}  = \boldsymbol{U}_{\rm MZI}\cdot\vec{R}_{\rm in}  + \vec{d}\,,\\
   & \nonumber \boldsymbol{\sigma}_{\rm out} = \boldsymbol{U}_{\rm MZI}\cdot\boldsymbol{\sigma}_{\rm in}\cdot \boldsymbol{U}_{\rm MZI}^{\dagger} + \boldsymbol{Y}\,,
\end{align}
where $\vec{d}$ is an added displacement and $\boldsymbol{Y}$ represents any added noise \cite{Sharma2022}. 
No displacements or additional noise are present inside our sensor; thus we set $\vec{d} = \vec{0}$ and $\boldsymbol{Y} = \boldsymbol{0}$.

Given an output state, we obtain a lower bound on the measurement sensitivity by calculating the quantum Fisher information of the state \cite{Oh2019, Safranek2019}:

\begin{equation}
    \label{eq:FisherInfo}
    F(\phi) = 2 \frac{\partial \vec{R}_{\rm out}^T}{\partial \phi}\cdot\boldsymbol{\sigma_{\rm out}}^{-1}\cdot\frac{\partial \vec{R}_{\rm out}}{\partial \phi} + \frac{1}{4} {\rm Tr}[(\frac{\partial \boldsymbol{\sigma_{\rm out}}}{\partial \phi}\cdot\boldsymbol{\sigma_{\rm out}}^{-1})^2].
\end{equation}
Consider the classical case, where the input is a coherent state of amplitude $|\alpha|$:
\begin{align}
    \label{eq:classicalMZI}
    \vec{R}_{\rm in} = 
    \begin{pmatrix}
         \sqrt{2}\alpha \\ 0 \\ 0 \\ 0
    \end{pmatrix}, \\
    \vec{R}_{\rm out} = 
    \begin{pmatrix}
         0 \\ \sqrt{2}\alpha \sin(\phi/2) \\ \sqrt{2}\alpha \cos(\phi/2) \\ 0
    \end{pmatrix}, \\
 \nonumber \boldsymbol{\sigma}_{\rm in} = \boldsymbol{I_4}\ = \boldsymbol{\sigma}_{\rm out}.
\end{align}
The Fisher information we obtain from this is $F = |\alpha|^2$. 
The information is linear in the photon number of the state $\bar{n} = |\alpha|^2$. 
In the quantum case, squeezed vacuum is also injected into the unused port with $\chi = 0$. 
The mean vector $\vec{R_{\rm out}}$ remains the same, but

\begin{equation}
    \label{squeezedmatrix}
    \boldsymbol{\sigma_{\rm in}} = 
    \begin{pmatrix}
        1 & 0 & 0 & 0 \\
        0 & 1 & 0 & 0 \\
        0 & 0 & e^{2r} & 0 \\
        0 & 0 & 0 & e^{-2r}
    \end{pmatrix},
\end{equation}
\begin{equation} \label{eq:Fsqueezed}
    F_{\rm squeezed} = |\alpha|^2 e^{2r} + \sinh{r}^2.
\end{equation}
This expression agrees with other methods of calculating the Fisher information for the same scenario \cite{Pezze2008,Ono2010}. 
In the limit where $|\alpha| \gg 1$, the second term in $F_{\rm squeezed}$ is negligible, and we can express the quantum advantage $Q$ in phase sensitivity from injecting squeezed light as:

\begin{equation}
    \label{eq:quantumadvantage}
    Q = \frac{\Delta^2\phi_{\rm classical}}{\Delta^2\phi_{\rm squeezed}} = \frac{F_{\rm squeezed}}{F_{\rm classical}} = e^{2r}.
\end{equation}
The term $e^{-2r}$ is the shot-noise reduction factor of a squeezed state along its squeezed quadrature. 
Thus, the maximal improvement of sensitivity given by this state is $e^{2r}$. 
State-of-the-art squeezing sources can achieve about an order of magnitude of quadrature squeezing, so we choose $r = 1$ for our simulations \cite{Mehmet2010, Vernon2019, Eberle2011}. 

\section{Loss and Noise}\label{appendix:B}

\begin{figure}[h]
    \centering
    \includegraphics[keepaspectratio, width=0.9\columnwidth]{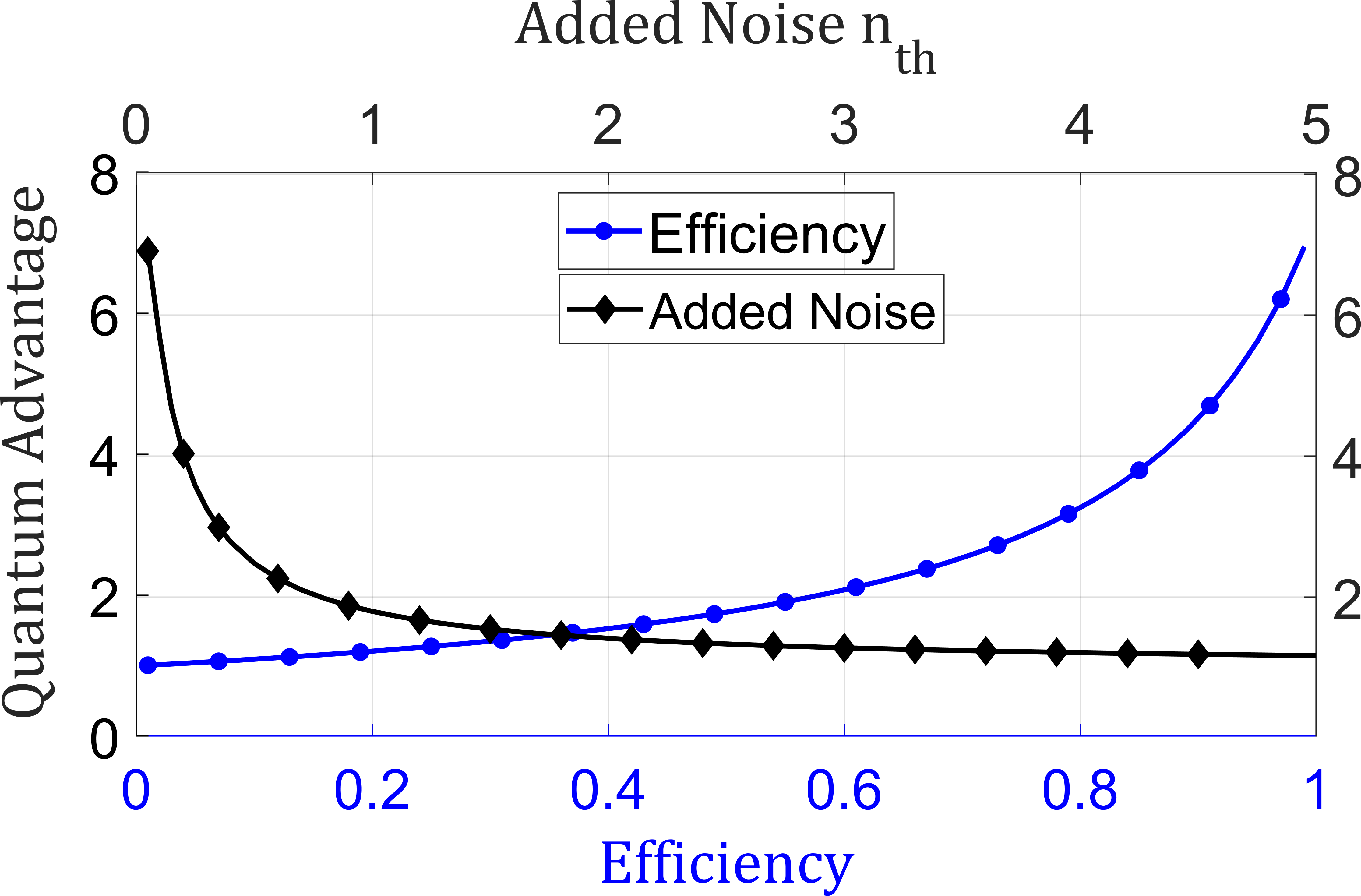}
    \caption{Results showing the degradation in quantum advantage of the sensor as one adds optical loss (blue, dotted), or thermal noise (black, diamond), described by an average number of thermal photons $n_th$. Loss effectively reduces the squeezing strength. Added thermal noise ruins the condition of having a shot-noise limited system, without which squeezing is ineffective.}
    \label{fig:figS2}
\end{figure}

The sensor does not infinitely scale to an arbitrarily high number of phases. 
For a meaningful quantum advantage, the number of input phase-locked squeezed pulses scales alongside $N$, which may become experimentally challenging. 
Additionally, dispersion, optical loss, and limited coherence will all set an upper limit on the number of optical elements that the squeezed pulses pass through. 
Dispersion and coherence can be engineered, but optical loss sets a harder limit.

By applying optical loss to all modes in the interferometer, we analyzed the extent to which the sensor can maintain a quantum advantage under lossy conditions. 
We implemented this with Eq. \eqref{eq:propagation} in which $\boldsymbol{U} = \sqrt{\epsilon}\boldsymbol{I}$, where $\boldsymbol{I}$ is the identity matrix and $\epsilon$ is the efficiency, $\vec{d} = \vec{0}$, and $\boldsymbol{Y} = (1-\epsilon)\boldsymbol{I}$.
Results are in Fig. \ref{fig:figS2}, which shows the same expected behavior from single-parameter sensing. 
A non-unity efficiency introduces added noise into the state, effectively reducing the squeezing strength and the resulting quantum advantage $Q$ \cite{Ono2010}.

Similarly, we analyzed how added thermal noise or phase noise affects $Q$. To implement added thermal noise, we directly added the matrix $n_{\rm th} \boldsymbol{I}$ with $n_{\rm th}$ photons to the output covariance matrix. Mixing the output state with a thermal state of average photon number $n_{th}$ reduces $Q$. 
As expected, when the sensor is no longer shot noise limited ($n_{th} > \bar{n}_{\rm squeezed}$), a large quantum advantage is no longer possible. 

\section{Numeric Scaling with Photon Number}\label{appendix:C}
\begin{figure}[h]
    \centering
    \includegraphics[keepaspectratio, width=0.9\columnwidth]{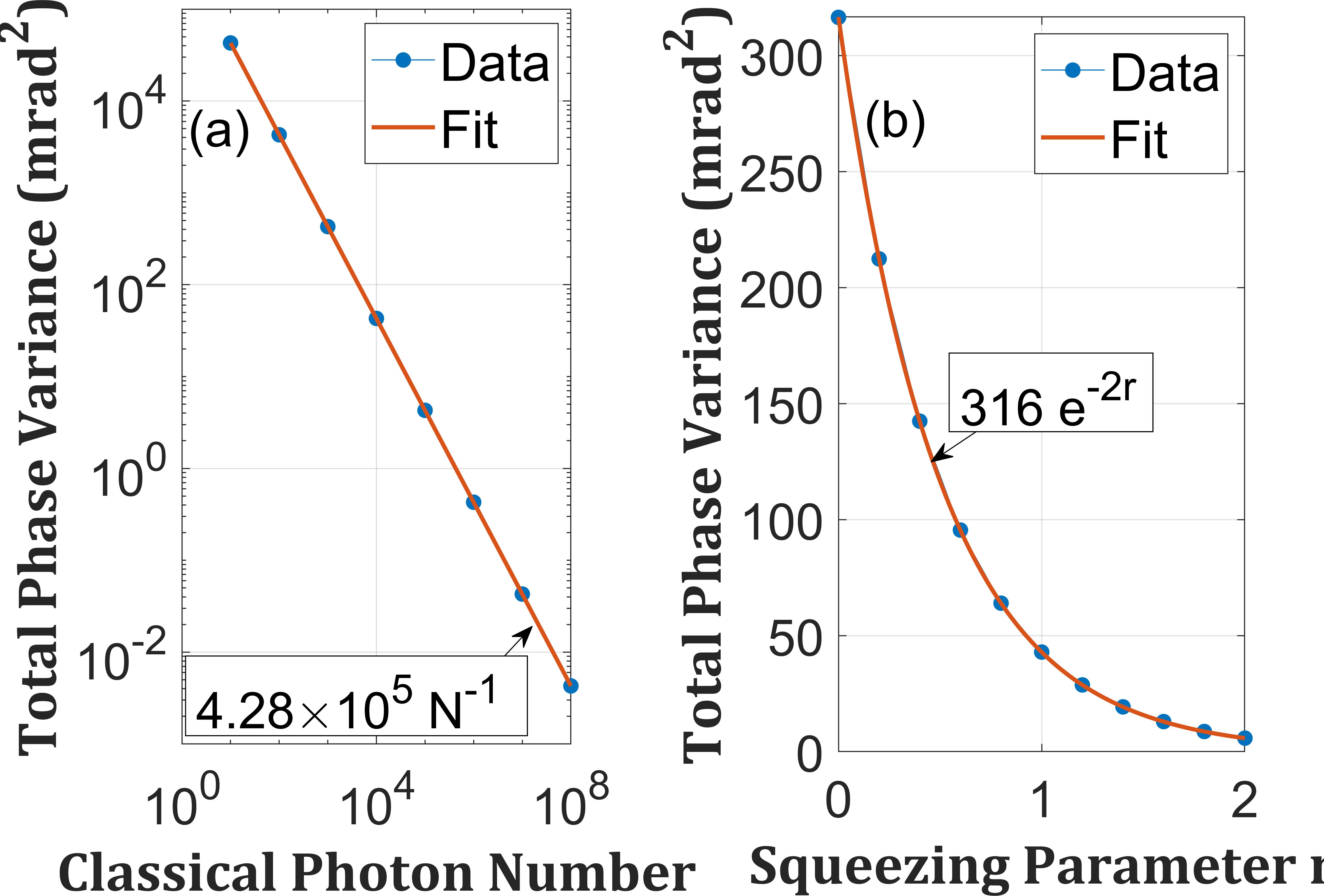}
    \caption{Simulation data verifying the scaling of the total phase sensitivity with respect to classical photon number (linear) and squeezing parameter (exponential). The fitted numerical constants are inset.}
    \label{fig:figS3}
\end{figure}

Fig. \ref{fig:figS3} verifies that our simulation results scale as expected with respect to classical photon number and the squeezing parameter. Data was acquired from the three-phase interferometer configuration as used in Fig. \ref{fig:3phDownsampled}. In Fig. \ref{fig:figS3}(a), the squeezing parameter $r$ was set to 1 while $|\alpha|^2$ varied from $10 - 10^8$. In Fig. \ref{fig:figS3}(b), $|\alpha|^2$ was set to $10^4$ while $r$ varied from $0-2$.

\section{Three-Phase Interferometer with Sequential Inputs}\label{appendix:D}
\begin{figure}[t]
    \centering
    \includegraphics[keepaspectratio, width=0.9\columnwidth]{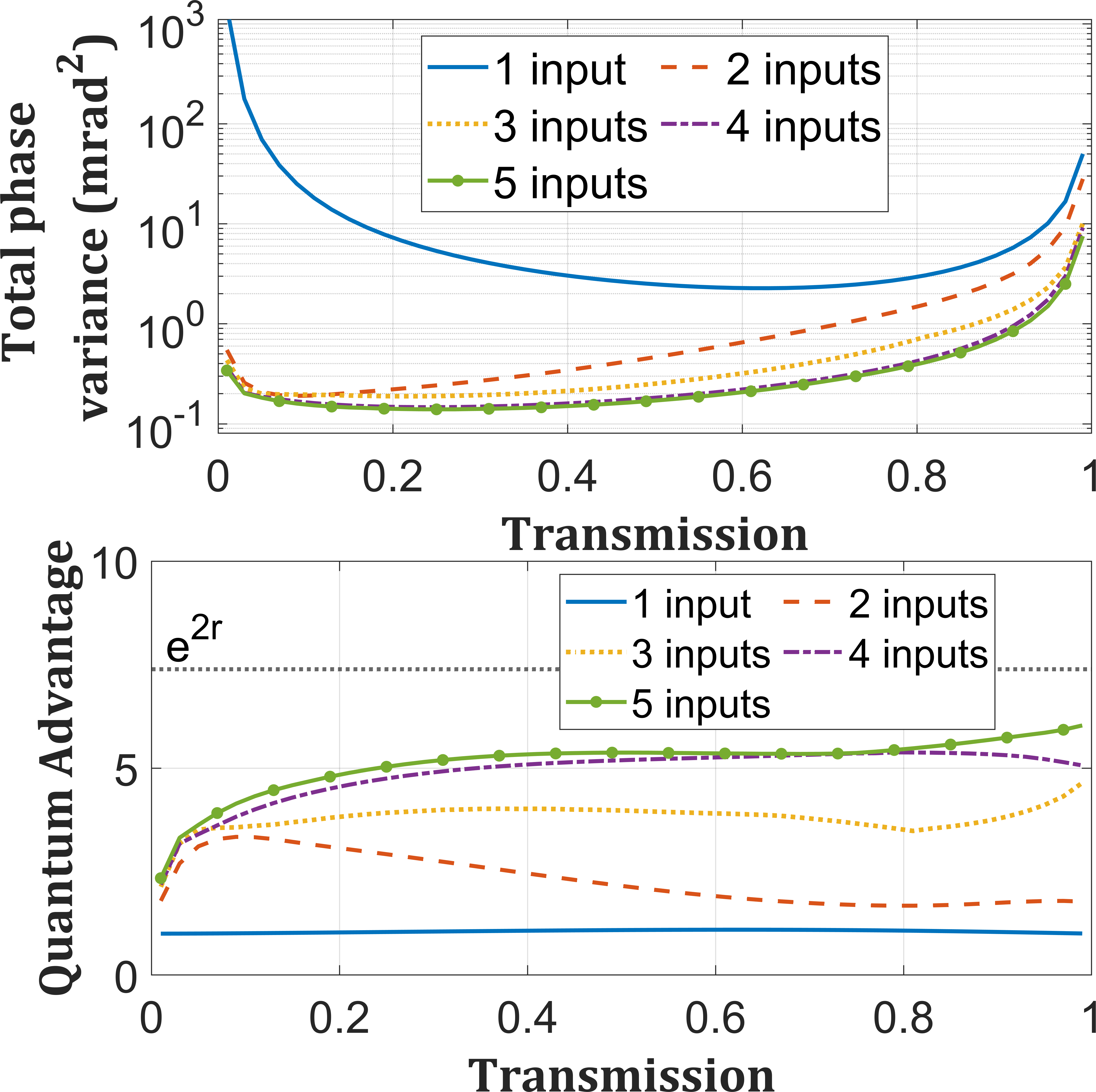}
    \caption{(a) Plot of the total phase sensitivity in a three-phase, bidirectional interferometer with up to five sequential displaced squeezed state inputs. The magnitude of $|\alpha|$ reduces with each added pulse to keep the total input photon number constant at $10^3$. (b) Plot of the quantum advantage $Q$ obtained from sequentially inputting up to five displaced squeezed states. The quantum advantage is shown to increase with more input pulses, asymptotically approaching the limit set by the amount of vacuum noise in the system.}
    \label{fig:3phRes}
\end{figure}

Another situation to consider in the same interferometer is sequential inputs of multiple displaced squeezed states, as in Fig. 3. 
Since continuous input of these states would wash out any timing information, we must assume that the sensor is initially unsqueezed. 
The additional power of many displaced inputs may provide an overall better sensitivity than the ``downsampled" case. 
Results are shown in Fig. \ref{fig:3phRes}. 
Similar to the analysis of the two-phase interferometer, we show the total phase variance and quantum advantage from using a varying number of inputs. 
A single input gives a $Q$ of at most 9.5\% at $T = 0.62$. 
However, with a second input, the light has one fewer interaction with vacuum noise, and so the sensor shows a stronger quantum advantage as well as a lower phase variance. 
A third pulse adds a smaller enhancement and more inputs show more diminishing returns on the sensitivity and quantum advantage. 
With more pulses comes greater ambiguity in which phase caused which output mode to have a phase shift, leading to relatively lower distinguishing information. 
Also, more pulses would require a longer measurement period, which is not accounted for in this analysis. 
Therefore, for simplicity in measurement, it may be more practical to maintain the squeezing inside the interferometer with the downsampled case. 
The phases can then be probed with maximal squeezing enhancement with either one or two displaced squeezed pulses.

\section{Multiparameter estimation advantage}\label{appendix:E}

\begin{figure}[H]
    \centering
    \includegraphics[keepaspectratio, width=0.9\columnwidth]{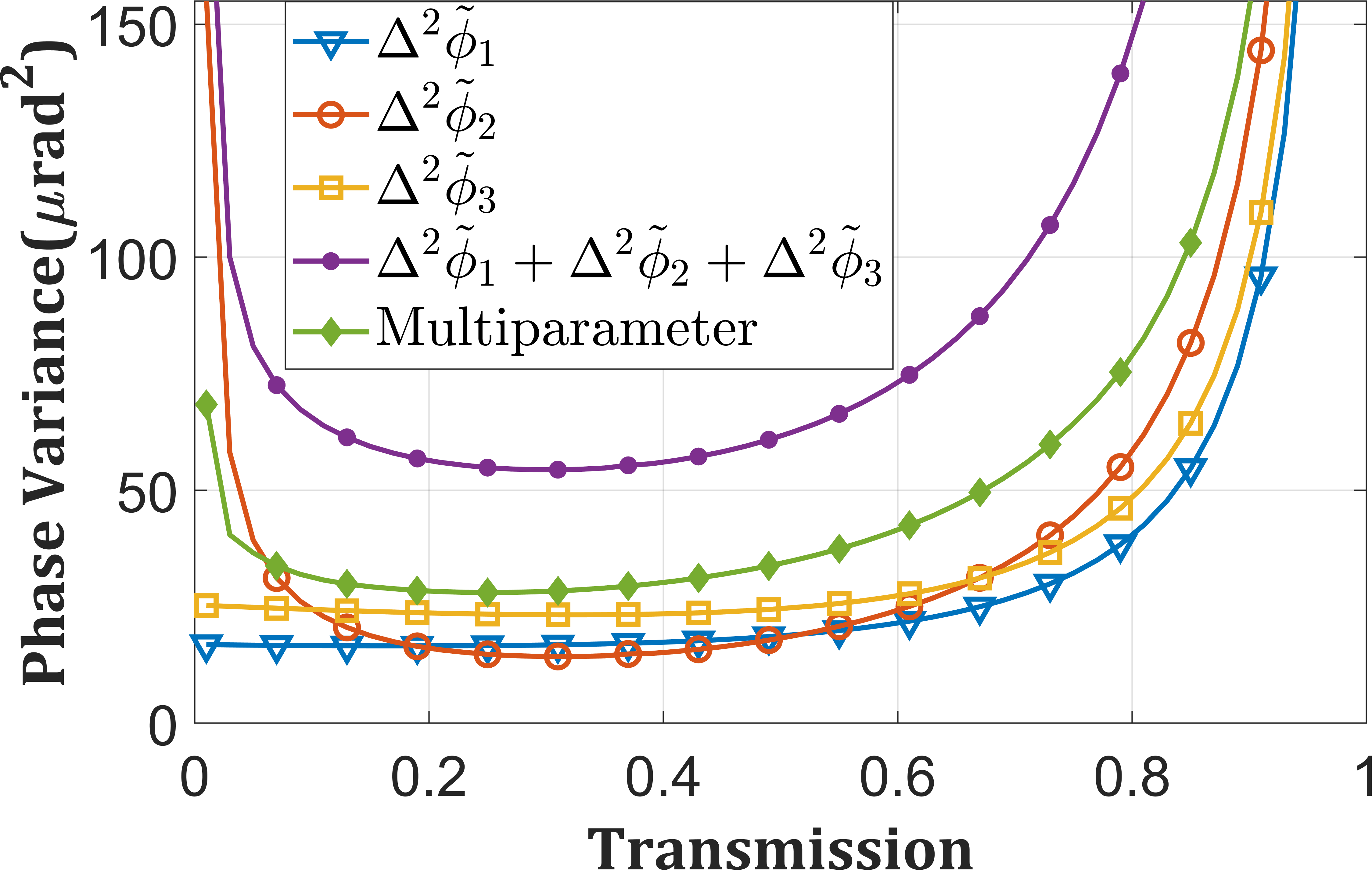}
    \caption{Comparison of measurement sensitivities from different methods in the three-phase interferometer, as a function of reflector transmission. The same sequential pulse procedure from Fig. \ref{fig:3phRes} is done here.  Traces for phases 1, 2, and 3 refer to single-parameter estimation and optimization of those phases, assuming the others are known. The sum of those traces as three separate measurements is compared with the total variance from measuring all phases at once in a multiparameter approach. The total phase variance from a multiparameter estimation is shown to be between 1.7 and 2.8 lower than that of sensing individual phases, showing the efficacy of sensing all parameters at once.}
    \label{fig:sensRes}
\end{figure}

Another figure of merit for a multiparameter sensor is its enhanced precision over simply measuring each parameter individually. Therefore, the multiparameter estimation from Fig. \ref{fig:3phRes} may also be compared with that of individual phase estimation in a series of three separate measurements.
Fig. \ref{fig:sensRes} shows three traces of the optimal variance for measuring each phase individually using five pulses, multiplied by three because each phase is measured one-third as often.
The sum of these variances is also plotted and compared to that of multiparameter estimation, where the total phase variance was optimized collectively.
The total variance of the multiparameter measurement is between $1.7$ and $2.8$ times lower than the sum of individual measurements.
This lower variance shows a separate advantage of using multiparameter estimation techniques, on top of using squeezed light. Other works predict a factor of $N$ improvement for using multiparameter estimation with $N$ parameters \cite{Humphreys2013, Ragy2016}, or a flat factor between 2 and 4 \cite{You2017, Gagatsos2016}, and so our maximum observed quantum advantage of 2.8 for three parameters is reasonable. A similar analysis on interferometers with more phases would be necessary to determine if the enhancement scales with $N$.

\section{Optimization}\label{appendix:F}

Optimization plays a key role in the physics of the sensor because it is not guaranteed that squeezing will reduce the shot noise in a homodyne-based measurement. 
A squeezed state of random squeezing angle would be equally likely to anti-squeeze the correct quadrature as it would to squeeze it. 
Thus, it is non-trivial that a simultaneous measurement on $N$ cascaded phases would have the correct squeezing angle for each output mode simultaneously. 
Our optimization algorithm ensures this as best it can given the geometry of the system. 
For an $N$-phase interferometer with $m$ input pulses, there will be $N + 2m - 1$ variables for the optimization (minus one because of a global phase). 
As an example, we return to the two-phase interferometer with two inputs. 
This has five variables: the measurable phases $\phi_1, \phi_2$, the input coherent state phase for each side $\theta_1 = 0, \theta_2$, and the squeezing angles for each side $\chi_1, \chi_2$. 

\begin{figure}[H]
    \centering
    \includegraphics[keepaspectratio, width=0.9\columnwidth]{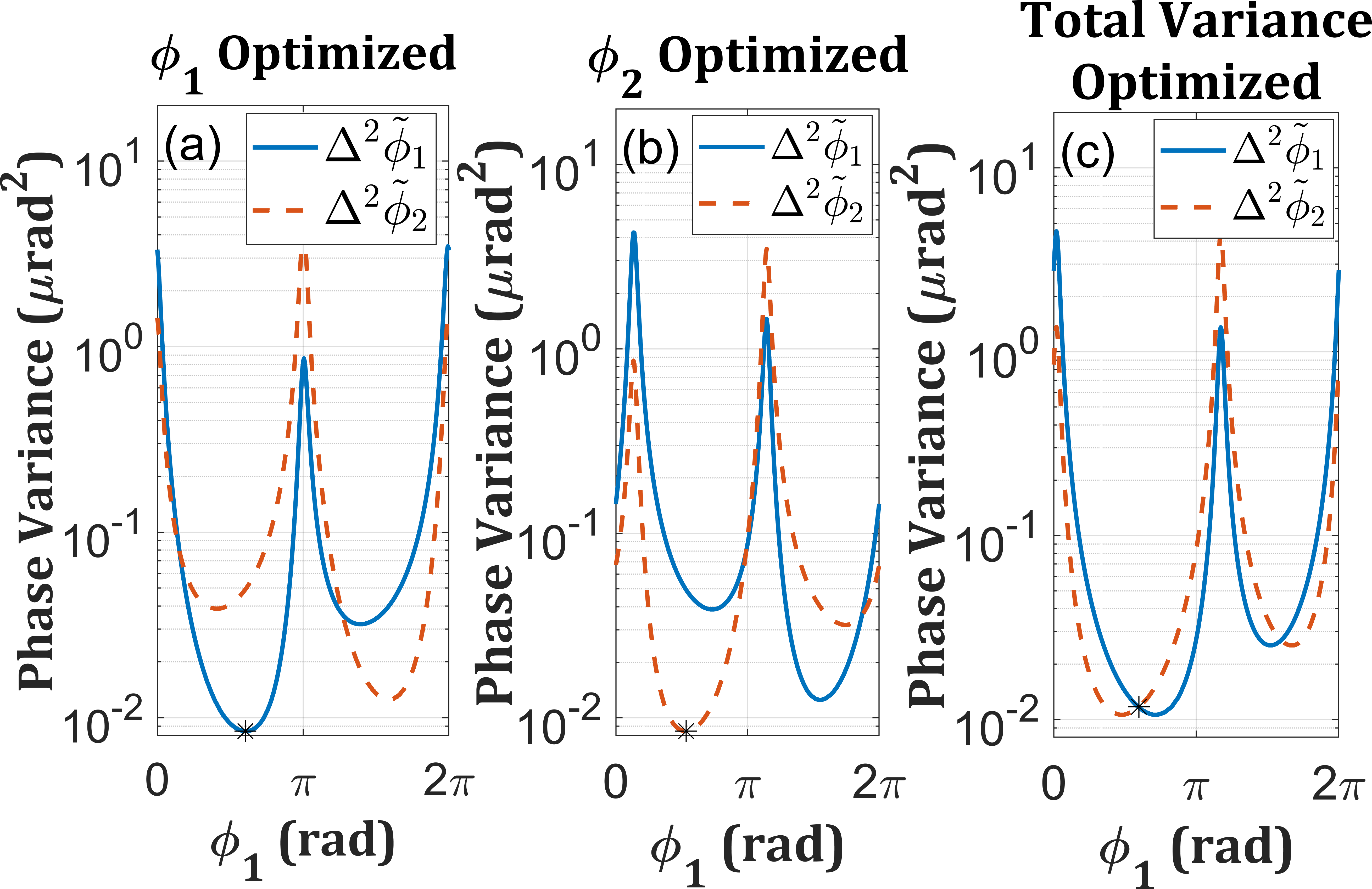}
    \caption{Comparison of optimized measurements with squeezing in the two-phase interferometer example and resulting phase variances as a function of $\phi_1$. As before, input coherent state amplitude $|\alpha| = 10^5$ and squeezing strength $r = 1$. We examined the point $T = 0.9$, where in Fig. 3 the full quantum advantage is relatively low. (a) A measurement optimized for $\phi_1$, where the input values were $\theta_1 = 0, \phi_2 = 2.89, \theta_2 = 3.39, \chi_1 = 1.91, \chi_2 = 3.63$. A full quantum advantage of $e^{2r}$ is achieved for measuring $\phi_1$ at $\phi_1 = 1.9$, but measuring $\phi_2$ at the same time only gives a quantum advantage of $1.27$. (b) A measurement optimized for measuring $\phi_2$, at the point $\theta_1 = 0, \phi_2 = 2.78, \theta_2 = 3.94, \chi_1 = 3.14, \chi_2 = 3.51$, showing the opposite effect where a measurement on $\phi_2$ gives a full $e^{2r}$ quantum advantage at $\phi_1 = 1.68$, at the expense of the sensitivity of the $\phi_1$ measurement. In (c), a compromise is reached at the point $\theta_1 = 0, \phi_2 = 2.65, \theta_2 = 3.93, \chi_1 = 1.87, \chi_2 = 3.44$. Measuring at this point with $\phi_1 = 1.88$ gives an equal, although slightly lower, quantum advantage for both phases.}
    \label{fig:2phase_optimization}
\end{figure}

Fig. \ref{fig:2phase_optimization} shows the results of three different optimizations on all but one variable, $\phi_1$, with the phase variances for each measurement plotted separately. 
In (a), $\Delta^2\phi_1$ is exclusively optimized, at the expense of the sensitivity of the $\phi_2$ measurement. 
In (b), $\Delta^2\phi_2$ is exclusively optimized, giving reversed results. 
In (c), the total phase variance from both measurements is optimized (as in Fig. 4), resulting in a compromise point where each measurement is close to optimal. 
The above comparison shows that it is indeed nontrivial to have multiple measurements simultaneously enhanced by squeezing, given the geometric limitations of the system.

\section{Appendix G: Fisher Information Diagonalization}\label{appendix:G}
The Fisher information matrices presented throughout this work have off-diagonal terms. It is possible to diagonalize these matrices, transforming into the eigenbasis of the Fisher information. Here we argue through example why diagonalization is not useful for a cascaded sensor. Let us take the Fisher information matrix from the two-phase sensor:

\begin{equation} \label{G1}
    \boldsymbol{F}(T) = 
    \begin{pmatrix}
        4-3T & T \\
        T & T
    \end{pmatrix} \,.
\end{equation}

Scaling with $|\alpha|^2$ and $r$ is implied, and not shown. Diagonalizing this Fisher information requires the matrix:

\begin{equation}\label{G2}
    D = 
    \begin{pmatrix}
        \frac{\epsilon_{+}}{T A_{+}} & \frac{\epsilon_{-}}{T A_{-}} \\
        A_{-}^{-1} & A_{+}^{-1}
    \end{pmatrix},
\end{equation}
where:
\begin{align}
    \epsilon_{\pm} = 2-2T \pm \sqrt{5T^2-8T+4},  \\  \nonumber
    A_{\pm} = \sqrt{1+\frac{\epsilon_{\pm}}{T}^2}.
\end{align}
The diagonalized Fisher information is:
\begin{align}
    F' = \begin{pmatrix}
        \lambda_{-} & 0 \\
        0 & \lambda_{+}
    \end{pmatrix}, \\
\lambda_{\pm} = 2-T \pm \sqrt{5T^2-8T+4}.    
\end{align}
The basis of $F'$ is no longer $\phi_1$ and $\phi_2$ but instead $D_{11} \phi_1 + D_{12} \phi_2$ and $D_{21} \phi_1 + D_{22} \phi_2$. Despite the basis change, we find that:
\begin{equation}
    Tr(F^{-1}) = Tr(F'^{-1}) = \frac{1}{T} + \frac{1}{2-2T},
\end{equation}
so there is no change in the total measurement variance. More importantly, $F'^{-1}$ represents covariance in a measurement basis that is not accessible or useful in a cascaded sensor. The entire point of this sensor layout is to have individually resolvable elements along a common path and matrix diagonalization destroys that premise by mixing parameters again.

\bibliography{mainbib}

\begin{thebibliography}{68}%
\makeatletter
\providecommand \@ifxundefined [1]{%
 \@ifx{#1\undefined}
}%
\providecommand \@ifnum [1]{%
 \ifnum #1\expandafter \@firstoftwo
 \else \expandafter \@secondoftwo
 \fi
}%
\providecommand \@ifx [1]{%
 \ifx #1\expandafter \@firstoftwo
 \else \expandafter \@secondoftwo
 \fi
}%
\providecommand \natexlab [1]{#1}%
\providecommand \enquote  [1]{``#1''}%
\providecommand \bibnamefont  [1]{#1}%
\providecommand \bibfnamefont [1]{#1}%
\providecommand \citenamefont [1]{#1}%
\providecommand \href@noop [0]{\@secondoftwo}%
\providecommand \href [0]{\begingroup \@sanitize@url \@href}%
\providecommand \@href[1]{\@@startlink{#1}\@@href}%
\providecommand \@@href[1]{\endgroup#1\@@endlink}%
\providecommand \@sanitize@url [0]{\catcode `\\12\catcode `\$12\catcode `\&12\catcode `\#12\catcode `\^12\catcode `\_12\catcode `\%12\relax}%
\providecommand \@@startlink[1]{}%
\providecommand \@@endlink[0]{}%
\providecommand \url  [0]{\begingroup\@sanitize@url \@url }%
\providecommand \@url [1]{\endgroup\@href {#1}{\urlprefix }}%
\providecommand \urlprefix  [0]{URL }%
\providecommand \Eprint [0]{\href }%
\providecommand \doibase [0]{https://doi.org/}%
\providecommand \selectlanguage [0]{\@gobble}%
\providecommand \bibinfo  [0]{\@secondoftwo}%
\providecommand \bibfield  [0]{\@secondoftwo}%
\providecommand \translation [1]{[#1]}%
\providecommand \BibitemOpen [0]{}%
\providecommand \bibitemStop [0]{}%
\providecommand \bibitemNoStop [0]{.\EOS\space}%
\providecommand \EOS [0]{\spacefactor3000\relax}%
\providecommand \BibitemShut  [1]{\csname bibitem#1\endcsname}%
\let\auto@bib@innerbib\@empty
\bibitem [{\citenamefont {Caves}(1981)}]{Caves1981}%
  \BibitemOpen
  \bibfield  {author} {\bibinfo {author} {\bibfnamefont {C.~M.}\ \bibnamefont {Caves}},\ }\href {https://doi.org/10.1103/PhysRevD.23.1693} {\bibfield  {journal} {\bibinfo  {journal} {Phys. Rev. D}\ }\textbf {\bibinfo {volume} {23}},\ \bibinfo {pages} {1693} (\bibinfo {year} {1981})}\BibitemShut {NoStop}%
\bibitem [{\citenamefont {Giovannetti}\ \emph {et~al.}(2004)\citenamefont {Giovannetti}, \citenamefont {Lloyd},\ and\ \citenamefont {Maccone}}]{Giovannetti2004}%
  \BibitemOpen
  \bibfield  {author} {\bibinfo {author} {\bibfnamefont {V.}~\bibnamefont {Giovannetti}}, \bibinfo {author} {\bibfnamefont {S.}~\bibnamefont {Lloyd}},\ and\ \bibinfo {author} {\bibfnamefont {L.}~\bibnamefont {Maccone}},\ }\href {https://doi.org/10.1126/science.1104149} {\bibfield  {journal} {\bibinfo  {journal} {Science}\ }\textbf {\bibinfo {volume} {306}},\ \bibinfo {pages} {1330} (\bibinfo {year} {2004})}\BibitemShut {NoStop}%
\bibitem [{\citenamefont {Dowling}\ and\ \citenamefont {Seshadreesan}(2015)}]{Dowling2015}%
  \BibitemOpen
  \bibfield  {author} {\bibinfo {author} {\bibfnamefont {J.~P.}\ \bibnamefont {Dowling}}\ and\ \bibinfo {author} {\bibfnamefont {K.~P.}\ \bibnamefont {Seshadreesan}},\ }\href {https://doi.org/10.1109/JLT.2014.2386795} {\bibfield  {journal} {\bibinfo  {journal} {Journal of Lightwave Technology}\ }\textbf {\bibinfo {volume} {33}},\ \bibinfo {pages} {2359} (\bibinfo {year} {2015})},\ \Eprint {https://arxiv.org/abs/1412.7578} {arXiv:1412.7578} \BibitemShut {NoStop}%
\bibitem [{\citenamefont {Bondurant}\ and\ \citenamefont {Shapiro}(1984)}]{Bondurant1984}%
  \BibitemOpen
  \bibfield  {author} {\bibinfo {author} {\bibfnamefont {R.~S.}\ \bibnamefont {Bondurant}}\ and\ \bibinfo {author} {\bibfnamefont {J.~H.}\ \bibnamefont {Shapiro}},\ }\href {https://doi.org/10.1103/PhysRevD.30.2548} {\bibfield  {journal} {\bibinfo  {journal} {Phys. Rev. D}\ }\textbf {\bibinfo {volume} {30}},\ \bibinfo {pages} {2548} (\bibinfo {year} {1984})}\BibitemShut {NoStop}%
\bibitem [{\citenamefont {Aasi}\ \emph {et~al.}(2013)\citenamefont {Aasi}, \citenamefont {Abadie}, \citenamefont {Abbott} \emph {et~al.}}]{Aasi2013}%
  \BibitemOpen
  \bibfield  {author} {\bibinfo {author} {\bibfnamefont {J.}~\bibnamefont {Aasi}}, \bibinfo {author} {\bibfnamefont {J.}~\bibnamefont {Abadie}}, \bibinfo {author} {\bibfnamefont {B.}~\bibnamefont {Abbott}}, \emph {et~al.} (\bibinfo {collaboration} {The LIGO Scientific Collaboration}),\ }\href {https://doi.org/10.1038/nphoton.2013.177} {\bibfield  {journal} {\bibinfo  {journal} {Nature Photonics}\ }\textbf {\bibinfo {volume} {7}},\ \bibinfo {pages} {613} (\bibinfo {year} {2013})},\ \Eprint {https://arxiv.org/abs/1310.0383} {arXiv:1310.0383} \BibitemShut {NoStop}%
\bibitem [{\citenamefont {Taylor}\ \emph {et~al.}(2013)\citenamefont {Taylor}, \citenamefont {Janousek}, \citenamefont {Daria}, \citenamefont {Knittel}, \citenamefont {Hage}, \citenamefont {Bachor},\ and\ \citenamefont {Bowen}}]{Taylor2013}%
  \BibitemOpen
  \bibfield  {author} {\bibinfo {author} {\bibfnamefont {M.~A.}\ \bibnamefont {Taylor}}, \bibinfo {author} {\bibfnamefont {J.}~\bibnamefont {Janousek}}, \bibinfo {author} {\bibfnamefont {V.}~\bibnamefont {Daria}}, \bibinfo {author} {\bibfnamefont {J.}~\bibnamefont {Knittel}}, \bibinfo {author} {\bibfnamefont {B.}~\bibnamefont {Hage}}, \bibinfo {author} {\bibfnamefont {H.-A.}\ \bibnamefont {Bachor}},\ and\ \bibinfo {author} {\bibfnamefont {W.~P.}\ \bibnamefont {Bowen}},\ }\href {https://doi.org/10.1038/nphoton.2012.346} {\bibfield  {journal} {\bibinfo  {journal} {Nature Photonics}\ }\textbf {\bibinfo {volume} {7}},\ \bibinfo {pages} {229} (\bibinfo {year} {2013})},\ \Eprint {https://arxiv.org/abs/1206.6928} {arXiv:1206.6928} \BibitemShut {NoStop}%
\bibitem [{\citenamefont {Ma}\ \emph {et~al.}(2011)\citenamefont {Ma}, \citenamefont {Wang}, \citenamefont {Sun},\ and\ \citenamefont {Nori}}]{MA201189}%
  \BibitemOpen
  \bibfield  {author} {\bibinfo {author} {\bibfnamefont {J.}~\bibnamefont {Ma}}, \bibinfo {author} {\bibfnamefont {X.}~\bibnamefont {Wang}}, \bibinfo {author} {\bibfnamefont {C.}~\bibnamefont {Sun}},\ and\ \bibinfo {author} {\bibfnamefont {F.}~\bibnamefont {Nori}},\ }\href {https://doi.org/https://doi.org/10.1016/j.physrep.2011.08.003} {\bibfield  {journal} {\bibinfo  {journal} {Physics Reports}\ }\textbf {\bibinfo {volume} {509}},\ \bibinfo {pages} {89} (\bibinfo {year} {2011})}\BibitemShut {NoStop}%
\bibitem [{\citenamefont {Mehmet}\ \emph {et~al.}(2011)\citenamefont {Mehmet}, \citenamefont {Ast}, \citenamefont {Eberle}, \citenamefont {Steinlechner}, \citenamefont {Vahlbruch},\ and\ \citenamefont {Schnabel}}]{Mehmet2011}%
  \BibitemOpen
  \bibfield  {author} {\bibinfo {author} {\bibfnamefont {M.}~\bibnamefont {Mehmet}}, \bibinfo {author} {\bibfnamefont {S.}~\bibnamefont {Ast}}, \bibinfo {author} {\bibfnamefont {T.}~\bibnamefont {Eberle}}, \bibinfo {author} {\bibfnamefont {S.}~\bibnamefont {Steinlechner}}, \bibinfo {author} {\bibfnamefont {H.}~\bibnamefont {Vahlbruch}},\ and\ \bibinfo {author} {\bibfnamefont {R.}~\bibnamefont {Schnabel}},\ }\href {https://doi.org/10.1364/OE.19.025763} {\bibfield  {journal} {\bibinfo  {journal} {Optics Express}\ }\textbf {\bibinfo {volume} {19}},\ \bibinfo {pages} {25763} (\bibinfo {year} {2011})},\ \Eprint {https://arxiv.org/abs/1110.3737} {arXiv:1110.3737} \BibitemShut {NoStop}%
\bibitem [{\citenamefont {Cox}\ \emph {et~al.}(2016)\citenamefont {Cox}, \citenamefont {Greve}, \citenamefont {Weiner},\ and\ \citenamefont {Thompson}}]{Cox2016}%
  \BibitemOpen
  \bibfield  {author} {\bibinfo {author} {\bibfnamefont {K.~C.}\ \bibnamefont {Cox}}, \bibinfo {author} {\bibfnamefont {G.~P.}\ \bibnamefont {Greve}}, \bibinfo {author} {\bibfnamefont {J.~M.}\ \bibnamefont {Weiner}},\ and\ \bibinfo {author} {\bibfnamefont {J.~K.}\ \bibnamefont {Thompson}},\ }\href {https://doi.org/10.1103/PhysRevLett.116.093602} {\bibfield  {journal} {\bibinfo  {journal} {Phys. Rev. Lett.}\ }\textbf {\bibinfo {volume} {116}},\ \bibinfo {pages} {093602} (\bibinfo {year} {2016})}\BibitemShut {NoStop}%
\bibitem [{\citenamefont {Crowley}\ \emph {et~al.}(2014)\citenamefont {Crowley}, \citenamefont {Datta}, \citenamefont {Barbieri},\ and\ \citenamefont {Walmsley}}]{Crowley2014}%
  \BibitemOpen
  \bibfield  {author} {\bibinfo {author} {\bibfnamefont {P.~J.~D.}\ \bibnamefont {Crowley}}, \bibinfo {author} {\bibfnamefont {A.}~\bibnamefont {Datta}}, \bibinfo {author} {\bibfnamefont {M.}~\bibnamefont {Barbieri}},\ and\ \bibinfo {author} {\bibfnamefont {I.~A.}\ \bibnamefont {Walmsley}},\ }\href {https://doi.org/10.1103/PhysRevA.89.023845} {\bibfield  {journal} {\bibinfo  {journal} {Physical Review A}\ }\textbf {\bibinfo {volume} {89}},\ \bibinfo {pages} {023845} (\bibinfo {year} {2014})},\ \Eprint {https://arxiv.org/abs/1206.0043} {arXiv:1206.0043} \BibitemShut {NoStop}%
\bibitem [{\citenamefont {Genoni}\ \emph {et~al.}(2011)\citenamefont {Genoni}, \citenamefont {Olivares},\ and\ \citenamefont {Paris}}]{Genoni2011}%
  \BibitemOpen
  \bibfield  {author} {\bibinfo {author} {\bibfnamefont {M.~G.}\ \bibnamefont {Genoni}}, \bibinfo {author} {\bibfnamefont {S.}~\bibnamefont {Olivares}},\ and\ \bibinfo {author} {\bibfnamefont {M.~G.~A.}\ \bibnamefont {Paris}},\ }\href {https://doi.org/10.1103/PhysRevLett.106.153603} {\bibfield  {journal} {\bibinfo  {journal} {Phys. Rev. Lett.}\ }\textbf {\bibinfo {volume} {106}},\ \bibinfo {pages} {153603} (\bibinfo {year} {2011})}\BibitemShut {NoStop}%
\bibitem [{\citenamefont {Vidrighin}\ \emph {et~al.}(2014)\citenamefont {Vidrighin}, \citenamefont {Donati}, \citenamefont {Genoni}, \citenamefont {Jin}, \citenamefont {Kolthammer}, \citenamefont {Kim}, \citenamefont {Datta}, \citenamefont {Barbieri},\ and\ \citenamefont {Walmsley}}]{Vidrighin2014}%
  \BibitemOpen
  \bibfield  {author} {\bibinfo {author} {\bibfnamefont {M.~D.}\ \bibnamefont {Vidrighin}}, \bibinfo {author} {\bibfnamefont {G.}~\bibnamefont {Donati}}, \bibinfo {author} {\bibfnamefont {M.~G.}\ \bibnamefont {Genoni}}, \bibinfo {author} {\bibfnamefont {X.-M.}\ \bibnamefont {Jin}}, \bibinfo {author} {\bibfnamefont {W.~S.}\ \bibnamefont {Kolthammer}}, \bibinfo {author} {\bibfnamefont {M.}~\bibnamefont {Kim}}, \bibinfo {author} {\bibfnamefont {A.}~\bibnamefont {Datta}}, \bibinfo {author} {\bibfnamefont {M.}~\bibnamefont {Barbieri}},\ and\ \bibinfo {author} {\bibfnamefont {I.~A.}\ \bibnamefont {Walmsley}},\ }\href {https://doi.org/10.1038/ncomms4532} {\bibfield  {journal} {\bibinfo  {journal} {Nature Communications}\ }\textbf {\bibinfo {volume} {5}},\ \bibinfo {pages} {3532} (\bibinfo {year} {2014})},\ \Eprint {https://arxiv.org/abs/1410.5353} {arXiv:1410.5353} \BibitemShut {NoStop}%
\bibitem [{\citenamefont {Li}\ \emph {et~al.}(2018)\citenamefont {Li}, \citenamefont {B{\'{i}}lek}, \citenamefont {Hoff}, \citenamefont {Madsen}, \citenamefont {Forstner}, \citenamefont {Prakash}, \citenamefont {Sch{\"{a}}fermeier}, \citenamefont {Gehring}, \citenamefont {Bowen},\ and\ \citenamefont {Andersen}}]{Li2018}%
  \BibitemOpen
  \bibfield  {author} {\bibinfo {author} {\bibfnamefont {B.-B.}\ \bibnamefont {Li}}, \bibinfo {author} {\bibfnamefont {J.}~\bibnamefont {B{\'{i}}lek}}, \bibinfo {author} {\bibfnamefont {U.~B.}\ \bibnamefont {Hoff}}, \bibinfo {author} {\bibfnamefont {L.~S.}\ \bibnamefont {Madsen}}, \bibinfo {author} {\bibfnamefont {S.}~\bibnamefont {Forstner}}, \bibinfo {author} {\bibfnamefont {V.}~\bibnamefont {Prakash}}, \bibinfo {author} {\bibfnamefont {C.}~\bibnamefont {Sch{\"{a}}fermeier}}, \bibinfo {author} {\bibfnamefont {T.}~\bibnamefont {Gehring}}, \bibinfo {author} {\bibfnamefont {W.~P.}\ \bibnamefont {Bowen}},\ and\ \bibinfo {author} {\bibfnamefont {U.~L.}\ \bibnamefont {Andersen}},\ }\href {https://doi.org/10.1364/OPTICA.5.000850} {\bibfield  {journal} {\bibinfo  {journal} {Optica}\ }\textbf {\bibinfo {volume} {5}},\ \bibinfo {pages} {850} (\bibinfo {year} {2018})}\BibitemShut {NoStop}%
\bibitem [{\citenamefont {Friel}\ \emph {et~al.}(2020)\citenamefont {Friel}, \citenamefont {Palittapongarnpim}, \citenamefont {Albarelli},\ and\ \citenamefont {Datta}}]{Friel2020}%
  \BibitemOpen
  \bibfield  {author} {\bibinfo {author} {\bibfnamefont {J.}~\bibnamefont {Friel}}, \bibinfo {author} {\bibfnamefont {P.}~\bibnamefont {Palittapongarnpim}}, \bibinfo {author} {\bibfnamefont {F.}~\bibnamefont {Albarelli}},\ and\ \bibinfo {author} {\bibfnamefont {A.}~\bibnamefont {Datta}},\ }\href {https://arxiv.org/abs/2008.01502} {\bibinfo {title} {Attainability of the holevo-cram\'er-rao bound for two-qubit 3d magnetometry}} (\bibinfo {year} {2020}),\ \Eprint {https://arxiv.org/abs/2008.01502} {arXiv:2008.01502 [quant-ph]} \BibitemShut {NoStop}%
\bibitem [{\citenamefont {Hou}\ \emph {et~al.}(2020)\citenamefont {Hou}, \citenamefont {Zhang}, \citenamefont {Xiang}, \citenamefont {Li}, \citenamefont {Guo}, \citenamefont {Chen}, \citenamefont {Liu},\ and\ \citenamefont {Yuan}}]{Hou2020}%
  \BibitemOpen
  \bibfield  {author} {\bibinfo {author} {\bibfnamefont {Z.}~\bibnamefont {Hou}}, \bibinfo {author} {\bibfnamefont {Z.}~\bibnamefont {Zhang}}, \bibinfo {author} {\bibfnamefont {G.-Y.}\ \bibnamefont {Xiang}}, \bibinfo {author} {\bibfnamefont {C.-F.}\ \bibnamefont {Li}}, \bibinfo {author} {\bibfnamefont {G.-C.}\ \bibnamefont {Guo}}, \bibinfo {author} {\bibfnamefont {H.}~\bibnamefont {Chen}}, \bibinfo {author} {\bibfnamefont {L.}~\bibnamefont {Liu}},\ and\ \bibinfo {author} {\bibfnamefont {H.}~\bibnamefont {Yuan}},\ }\href {https://doi.org/10.1103/PhysRevLett.125.020501} {\bibfield  {journal} {\bibinfo  {journal} {Phys. Rev. Lett.}\ }\textbf {\bibinfo {volume} {125}},\ \bibinfo {pages} {020501} (\bibinfo {year} {2020})}\BibitemShut {NoStop}%
\bibitem [{\citenamefont {K{\'{o}}m{\'{a}}r}\ \emph {et~al.}(2014)\citenamefont {K{\'{o}}m{\'{a}}r}, \citenamefont {Kessler}, \citenamefont {Bishof}, \citenamefont {Jiang}, \citenamefont {S{\o}rensen}, \citenamefont {Ye},\ and\ \citenamefont {Lukin}}]{Komar2014}%
  \BibitemOpen
  \bibfield  {author} {\bibinfo {author} {\bibfnamefont {P.}~\bibnamefont {K{\'{o}}m{\'{a}}r}}, \bibinfo {author} {\bibfnamefont {E.~M.}\ \bibnamefont {Kessler}}, \bibinfo {author} {\bibfnamefont {M.}~\bibnamefont {Bishof}}, \bibinfo {author} {\bibfnamefont {L.}~\bibnamefont {Jiang}}, \bibinfo {author} {\bibfnamefont {A.~S.}\ \bibnamefont {S{\o}rensen}}, \bibinfo {author} {\bibfnamefont {J.}~\bibnamefont {Ye}},\ and\ \bibinfo {author} {\bibfnamefont {M.~D.}\ \bibnamefont {Lukin}},\ }\href {https://doi.org/10.1038/nphys3000} {\bibfield  {journal} {\bibinfo  {journal} {Nature Physics}\ }\textbf {\bibinfo {volume} {10}},\ \bibinfo {pages} {582} (\bibinfo {year} {2014})},\ \Eprint {https://arxiv.org/abs/1310.6045} {arXiv:1310.6045} \BibitemShut {NoStop}%
\bibitem [{\citenamefont {Suzuki}\ \emph {et~al.}(2020)\citenamefont {Suzuki}, \citenamefont {Yang},\ and\ \citenamefont {Hayashi}}]{Suzuki2020}%
  \BibitemOpen
  \bibfield  {author} {\bibinfo {author} {\bibfnamefont {J.}~\bibnamefont {Suzuki}}, \bibinfo {author} {\bibfnamefont {Y.}~\bibnamefont {Yang}},\ and\ \bibinfo {author} {\bibfnamefont {M.}~\bibnamefont {Hayashi}},\ }\href {https://doi.org/10.1088/1751-8121/ab8b78} {\bibfield  {journal} {\bibinfo  {journal} {Journal of Physics A: Mathematical and Theoretical}\ }\textbf {\bibinfo {volume} {53}},\ \bibinfo {pages} {453001} (\bibinfo {year} {2020})},\ \Eprint {https://arxiv.org/abs/1911.02790} {arXiv:1911.02790} \BibitemShut {NoStop}%
\bibitem [{\citenamefont {Bisketzi}\ \emph {et~al.}(2019)\citenamefont {Bisketzi}, \citenamefont {Branford},\ and\ \citenamefont {Datta}}]{Bisketzi2019}%
  \BibitemOpen
  \bibfield  {author} {\bibinfo {author} {\bibfnamefont {E.}~\bibnamefont {Bisketzi}}, \bibinfo {author} {\bibfnamefont {D.}~\bibnamefont {Branford}},\ and\ \bibinfo {author} {\bibfnamefont {A.}~\bibnamefont {Datta}},\ }\href {https://doi.org/10.1088/1367-2630/ab58a0} {\bibfield  {journal} {\bibinfo  {journal} {New Journal of Physics}\ }\textbf {\bibinfo {volume} {21}},\ \bibinfo {pages} {123032} (\bibinfo {year} {2019})},\ \Eprint {https://arxiv.org/abs/1907.11657} {arXiv:1907.11657} \BibitemShut {NoStop}%
\bibitem [{\citenamefont {Albarelli}\ \emph {et~al.}(2020)\citenamefont {Albarelli}, \citenamefont {Barbieri}, \citenamefont {Genoni},\ and\ \citenamefont {Gianani}}]{Albarelli2020}%
  \BibitemOpen
  \bibfield  {author} {\bibinfo {author} {\bibfnamefont {F.}~\bibnamefont {Albarelli}}, \bibinfo {author} {\bibfnamefont {M.}~\bibnamefont {Barbieri}}, \bibinfo {author} {\bibfnamefont {M.}~\bibnamefont {Genoni}},\ and\ \bibinfo {author} {\bibfnamefont {I.}~\bibnamefont {Gianani}},\ }\href {https://doi.org/10.1016/j.physleta.2020.126311} {\bibfield  {journal} {\bibinfo  {journal} {Physics Letters A}\ }\textbf {\bibinfo {volume} {384}},\ \bibinfo {pages} {126311} (\bibinfo {year} {2020})},\ \Eprint {https://arxiv.org/abs/1911.12067} {arXiv:1911.12067} \BibitemShut {NoStop}%
\bibitem [{\citenamefont {Fiderer}\ \emph {et~al.}(2021)\citenamefont {Fiderer}, \citenamefont {Tufarelli}, \citenamefont {Piano},\ and\ \citenamefont {Adesso}}]{Fiderer2021}%
  \BibitemOpen
  \bibfield  {author} {\bibinfo {author} {\bibfnamefont {L.~J.}\ \bibnamefont {Fiderer}}, \bibinfo {author} {\bibfnamefont {T.}~\bibnamefont {Tufarelli}}, \bibinfo {author} {\bibfnamefont {S.}~\bibnamefont {Piano}},\ and\ \bibinfo {author} {\bibfnamefont {G.}~\bibnamefont {Adesso}},\ }\href {https://doi.org/10.1103/PRXQuantum.2.020308} {\bibfield  {journal} {\bibinfo  {journal} {PRX Quantum}\ }\textbf {\bibinfo {volume} {2}},\ \bibinfo {pages} {020308} (\bibinfo {year} {2021})},\ \Eprint {https://arxiv.org/abs/2012.01572} {arXiv:2012.01572} \BibitemShut {NoStop}%
\bibitem [{\citenamefont {Ragy}\ \emph {et~al.}(2016)\citenamefont {Ragy}, \citenamefont {Jarzyna},\ and\ \citenamefont {Demkowicz-Dobrza\ifmmode~\acute{n}\else \'{n}\fi{}ski}}]{Ragy2016}%
  \BibitemOpen
  \bibfield  {author} {\bibinfo {author} {\bibfnamefont {S.}~\bibnamefont {Ragy}}, \bibinfo {author} {\bibfnamefont {M.}~\bibnamefont {Jarzyna}},\ and\ \bibinfo {author} {\bibfnamefont {R.}~\bibnamefont {Demkowicz-Dobrza\ifmmode~\acute{n}\else \'{n}\fi{}ski}},\ }\href {https://doi.org/10.1103/PhysRevA.94.052108} {\bibfield  {journal} {\bibinfo  {journal} {Phys. Rev. A}\ }\textbf {\bibinfo {volume} {94}},\ \bibinfo {pages} {052108} (\bibinfo {year} {2016})}\BibitemShut {NoStop}%
\bibitem [{\citenamefont {Szczykulska}\ \emph {et~al.}(2016)\citenamefont {Szczykulska}, \citenamefont {Baumgratz},\ and\ \citenamefont {Datta}}]{Szczykulska2016}%
  \BibitemOpen
  \bibfield  {author} {\bibinfo {author} {\bibfnamefont {M.}~\bibnamefont {Szczykulska}}, \bibinfo {author} {\bibfnamefont {T.}~\bibnamefont {Baumgratz}},\ and\ \bibinfo {author} {\bibfnamefont {A.}~\bibnamefont {Datta}},\ }\href {https://doi.org/10.1080/23746149.2016.1230476} {\bibfield  {journal} {\bibinfo  {journal} {Advances in Physics: X}\ }\textbf {\bibinfo {volume} {1}},\ \bibinfo {pages} {621} (\bibinfo {year} {2016})},\ \Eprint {https://arxiv.org/abs/1604.02615} {arXiv:1604.02615} \BibitemShut {NoStop}%
\bibitem [{\citenamefont {Albarelli}\ and\ \citenamefont {Demkowicz-Dobrza{\'{n}}ski}(2022)}]{Albarelli2022}%
  \BibitemOpen
  \bibfield  {author} {\bibinfo {author} {\bibfnamefont {F.}~\bibnamefont {Albarelli}}\ and\ \bibinfo {author} {\bibfnamefont {R.}~\bibnamefont {Demkowicz-Dobrza{\'{n}}ski}},\ }\href {https://doi.org/10.1103/PhysRevX.12.011039} {\bibfield  {journal} {\bibinfo  {journal} {Physical Review X}\ }\textbf {\bibinfo {volume} {12}},\ \bibinfo {pages} {011039} (\bibinfo {year} {2022})}\BibitemShut {NoStop}%
\bibitem [{\citenamefont {Gagatsos}\ \emph {et~al.}(2016)\citenamefont {Gagatsos}, \citenamefont {Branford},\ and\ \citenamefont {Datta}}]{Gagatsos2016}%
  \BibitemOpen
  \bibfield  {author} {\bibinfo {author} {\bibfnamefont {C.~N.}\ \bibnamefont {Gagatsos}}, \bibinfo {author} {\bibfnamefont {D.}~\bibnamefont {Branford}},\ and\ \bibinfo {author} {\bibfnamefont {A.}~\bibnamefont {Datta}},\ }\href {https://doi.org/10.1103/PhysRevA.94.042342} {\bibfield  {journal} {\bibinfo  {journal} {Physical Review A}\ }\textbf {\bibinfo {volume} {94}},\ \bibinfo {pages} {042342} (\bibinfo {year} {2016})},\ \Eprint {https://arxiv.org/abs/1605.04819} {arXiv:1605.04819} \BibitemShut {NoStop}%
\bibitem [{\citenamefont {You}\ \emph {et~al.}(2017)\citenamefont {You}, \citenamefont {Adhikari}, \citenamefont {Chi}, \citenamefont {LaBorde}, \citenamefont {Matyas}, \citenamefont {Zhang}, \citenamefont {Su}, \citenamefont {Byrnes}, \citenamefont {Lu}, \citenamefont {Dowling},\ and\ \citenamefont {Olson}}]{You2017}%
  \BibitemOpen
  \bibfield  {author} {\bibinfo {author} {\bibfnamefont {C.}~\bibnamefont {You}}, \bibinfo {author} {\bibfnamefont {S.}~\bibnamefont {Adhikari}}, \bibinfo {author} {\bibfnamefont {Y.}~\bibnamefont {Chi}}, \bibinfo {author} {\bibfnamefont {M.~L.}\ \bibnamefont {LaBorde}}, \bibinfo {author} {\bibfnamefont {C.~T.}\ \bibnamefont {Matyas}}, \bibinfo {author} {\bibfnamefont {C.}~\bibnamefont {Zhang}}, \bibinfo {author} {\bibfnamefont {Z.}~\bibnamefont {Su}}, \bibinfo {author} {\bibfnamefont {T.}~\bibnamefont {Byrnes}}, \bibinfo {author} {\bibfnamefont {C.}~\bibnamefont {Lu}}, \bibinfo {author} {\bibfnamefont {J.~P.}\ \bibnamefont {Dowling}},\ and\ \bibinfo {author} {\bibfnamefont {J.~P.}\ \bibnamefont {Olson}},\ }\href {https://doi.org/10.1088/2040-8986/aa9133} {\bibfield  {journal} {\bibinfo  {journal} {Journal of Optics}\ }\textbf {\bibinfo {volume} {19}},\ \bibinfo {pages} {124002} (\bibinfo {year} {2017})}\BibitemShut {NoStop}%
\bibitem [{\citenamefont {G\'orecki}\ and\ \citenamefont {Demkowicz-Dobrza\ifmmode~\acute{n}\else \'{n}\fi{}ski}(2022)}]{Gorecki2022}%
  \BibitemOpen
  \bibfield  {author} {\bibinfo {author} {\bibfnamefont {W.}~\bibnamefont {G\'orecki}}\ and\ \bibinfo {author} {\bibfnamefont {R.}~\bibnamefont {Demkowicz-Dobrza\ifmmode~\acute{n}\else \'{n}\fi{}ski}},\ }\href {https://doi.org/10.1103/PhysRevLett.128.040504} {\bibfield  {journal} {\bibinfo  {journal} {Phys. Rev. Lett.}\ }\textbf {\bibinfo {volume} {128}},\ \bibinfo {pages} {040504} (\bibinfo {year} {2022})}\BibitemShut {NoStop}%
\bibitem [{\citenamefont {Humphreys}\ \emph {et~al.}(2013)\citenamefont {Humphreys}, \citenamefont {Barbieri}, \citenamefont {Datta},\ and\ \citenamefont {Walmsley}}]{Humphreys2013}%
  \BibitemOpen
  \bibfield  {author} {\bibinfo {author} {\bibfnamefont {P.~C.}\ \bibnamefont {Humphreys}}, \bibinfo {author} {\bibfnamefont {M.}~\bibnamefont {Barbieri}}, \bibinfo {author} {\bibfnamefont {A.}~\bibnamefont {Datta}},\ and\ \bibinfo {author} {\bibfnamefont {I.~A.}\ \bibnamefont {Walmsley}},\ }\href {https://doi.org/10.1103/PhysRevLett.111.070403} {\bibfield  {journal} {\bibinfo  {journal} {Physical Review Letters}\ }\textbf {\bibinfo {volume} {111}},\ \bibinfo {pages} {070403} (\bibinfo {year} {2013})},\ \Eprint {https://arxiv.org/abs/1307.7653} {arXiv:1307.7653} \BibitemShut {NoStop}%
\bibitem [{\citenamefont {Polino}\ \emph {et~al.}(2019)\citenamefont {Polino}, \citenamefont {Riva}, \citenamefont {Valeri}, \citenamefont {Silvestri}, \citenamefont {Corrielli}, \citenamefont {Crespi}, \citenamefont {Spagnolo}, \citenamefont {Osellame},\ and\ \citenamefont {Sciarrino}}]{Polino2019}%
  \BibitemOpen
  \bibfield  {author} {\bibinfo {author} {\bibfnamefont {E.}~\bibnamefont {Polino}}, \bibinfo {author} {\bibfnamefont {M.}~\bibnamefont {Riva}}, \bibinfo {author} {\bibfnamefont {M.}~\bibnamefont {Valeri}}, \bibinfo {author} {\bibfnamefont {R.}~\bibnamefont {Silvestri}}, \bibinfo {author} {\bibfnamefont {G.}~\bibnamefont {Corrielli}}, \bibinfo {author} {\bibfnamefont {A.}~\bibnamefont {Crespi}}, \bibinfo {author} {\bibfnamefont {N.}~\bibnamefont {Spagnolo}}, \bibinfo {author} {\bibfnamefont {R.}~\bibnamefont {Osellame}},\ and\ \bibinfo {author} {\bibfnamefont {F.}~\bibnamefont {Sciarrino}},\ }\href {https://doi.org/10.1364/OPTICA.6.000288} {\bibfield  {journal} {\bibinfo  {journal} {Optica}\ }\textbf {\bibinfo {volume} {6}},\ \bibinfo {pages} {288} (\bibinfo {year} {2019})}\BibitemShut {NoStop}%
\bibitem [{\citenamefont {Zhang}\ and\ \citenamefont {Chan}(2017)}]{Zhang2017}%
  \BibitemOpen
  \bibfield  {author} {\bibinfo {author} {\bibfnamefont {L.}~\bibnamefont {Zhang}}\ and\ \bibinfo {author} {\bibfnamefont {K.~W.~C.}\ \bibnamefont {Chan}},\ }\href {https://doi.org/10.1103/PhysRevA.95.032321} {\bibfield  {journal} {\bibinfo  {journal} {Physical Review A}\ }\textbf {\bibinfo {volume} {95}},\ \bibinfo {pages} {032321} (\bibinfo {year} {2017})},\ \Eprint {https://arxiv.org/abs/1703.00063} {arXiv:1703.00063} \BibitemShut {NoStop}%
\bibitem [{\citenamefont {Malitesta}\ \emph {et~al.}(2023)\citenamefont {Malitesta}, \citenamefont {Smerzi},\ and\ \citenamefont {Pezz{\`{e}}}}]{Malitesta2023}%
  \BibitemOpen
  \bibfield  {author} {\bibinfo {author} {\bibfnamefont {M.}~\bibnamefont {Malitesta}}, \bibinfo {author} {\bibfnamefont {A.}~\bibnamefont {Smerzi}},\ and\ \bibinfo {author} {\bibfnamefont {L.}~\bibnamefont {Pezz{\`{e}}}},\ }\href {https://doi.org/10.1103/PhysRevA.108.032621} {\bibfield  {journal} {\bibinfo  {journal} {Physical Review A}\ }\textbf {\bibinfo {volume} {108}},\ \bibinfo {pages} {032621} (\bibinfo {year} {2023})},\ \Eprint {https://arxiv.org/abs/2109.09178} {arXiv:2109.09178} \BibitemShut {NoStop}%
\bibitem [{\citenamefont {Valeri}\ \emph {et~al.}(2023)\citenamefont {Valeri}, \citenamefont {Cimini}, \citenamefont {Piacentini}, \citenamefont {Ceccarelli}, \citenamefont {Polino}, \citenamefont {Hoch}, \citenamefont {Bizzarri}, \citenamefont {Corrielli}, \citenamefont {Spagnolo}, \citenamefont {Osellame},\ and\ \citenamefont {Sciarrino}}]{Valeri2023}%
  \BibitemOpen
  \bibfield  {author} {\bibinfo {author} {\bibfnamefont {M.}~\bibnamefont {Valeri}}, \bibinfo {author} {\bibfnamefont {V.}~\bibnamefont {Cimini}}, \bibinfo {author} {\bibfnamefont {S.}~\bibnamefont {Piacentini}}, \bibinfo {author} {\bibfnamefont {F.}~\bibnamefont {Ceccarelli}}, \bibinfo {author} {\bibfnamefont {E.}~\bibnamefont {Polino}}, \bibinfo {author} {\bibfnamefont {F.}~\bibnamefont {Hoch}}, \bibinfo {author} {\bibfnamefont {G.}~\bibnamefont {Bizzarri}}, \bibinfo {author} {\bibfnamefont {G.}~\bibnamefont {Corrielli}}, \bibinfo {author} {\bibfnamefont {N.}~\bibnamefont {Spagnolo}}, \bibinfo {author} {\bibfnamefont {R.}~\bibnamefont {Osellame}},\ and\ \bibinfo {author} {\bibfnamefont {F.}~\bibnamefont {Sciarrino}},\ }\href {https://doi.org/10.1103/PhysRevResearch.5.013138} {\bibfield  {journal} {\bibinfo  {journal} {Physical Review Research}\ }\textbf {\bibinfo {volume} {5}},\ \bibinfo {pages} {013138} (\bibinfo {year} {2023})},\ \Eprint {https://arxiv.org/abs/2208.14473} {arXiv:2208.14473} \BibitemShut
  {NoStop}%
\bibitem [{\citenamefont {Proctor}\ \emph {et~al.}(2018)\citenamefont {Proctor}, \citenamefont {Knott},\ and\ \citenamefont {Dunningham}}]{Proctor2018}%
  \BibitemOpen
  \bibfield  {author} {\bibinfo {author} {\bibfnamefont {T.~J.}\ \bibnamefont {Proctor}}, \bibinfo {author} {\bibfnamefont {P.~A.}\ \bibnamefont {Knott}},\ and\ \bibinfo {author} {\bibfnamefont {J.~A.}\ \bibnamefont {Dunningham}},\ }\href {https://doi.org/10.1103/PhysRevLett.120.080501} {\bibfield  {journal} {\bibinfo  {journal} {Phys. Rev. Lett.}\ }\textbf {\bibinfo {volume} {120}},\ \bibinfo {pages} {080501} (\bibinfo {year} {2018})}\BibitemShut {NoStop}%
\bibitem [{\citenamefont {Oh}\ \emph {et~al.}(2020)\citenamefont {Oh}, \citenamefont {Lee}, \citenamefont {Lie},\ and\ \citenamefont {Jeong}}]{Oh2020}%
  \BibitemOpen
  \bibfield  {author} {\bibinfo {author} {\bibfnamefont {C.}~\bibnamefont {Oh}}, \bibinfo {author} {\bibfnamefont {C.}~\bibnamefont {Lee}}, \bibinfo {author} {\bibfnamefont {S.~H.}\ \bibnamefont {Lie}},\ and\ \bibinfo {author} {\bibfnamefont {H.}~\bibnamefont {Jeong}},\ }\href {https://doi.org/10.1103/PhysRevResearch.2.023030} {\bibfield  {journal} {\bibinfo  {journal} {Physical Review Research}\ }\textbf {\bibinfo {volume} {2}},\ \bibinfo {pages} {023030} (\bibinfo {year} {2020})},\ \Eprint {https://arxiv.org/abs/1910.00823} {arXiv:1910.00823} \BibitemShut {NoStop}%
\bibitem [{\citenamefont {Guo}\ \emph {et~al.}(2020)\citenamefont {Guo}, \citenamefont {Breum}, \citenamefont {Borregaard}, \citenamefont {Izumi}, \citenamefont {Larsen}, \citenamefont {Gehring}, \citenamefont {Christandl}, \citenamefont {Neergaard-Nielsen},\ and\ \citenamefont {Andersen}}]{Guo2020}%
  \BibitemOpen
  \bibfield  {author} {\bibinfo {author} {\bibfnamefont {X.}~\bibnamefont {Guo}}, \bibinfo {author} {\bibfnamefont {C.~R.}\ \bibnamefont {Breum}}, \bibinfo {author} {\bibfnamefont {J.}~\bibnamefont {Borregaard}}, \bibinfo {author} {\bibfnamefont {S.}~\bibnamefont {Izumi}}, \bibinfo {author} {\bibfnamefont {M.~V.}\ \bibnamefont {Larsen}}, \bibinfo {author} {\bibfnamefont {T.}~\bibnamefont {Gehring}}, \bibinfo {author} {\bibfnamefont {M.}~\bibnamefont {Christandl}}, \bibinfo {author} {\bibfnamefont {J.~S.}\ \bibnamefont {Neergaard-Nielsen}},\ and\ \bibinfo {author} {\bibfnamefont {U.~L.}\ \bibnamefont {Andersen}},\ }\href {https://doi.org/10.1038/s41567-019-0743-x} {\bibfield  {journal} {\bibinfo  {journal} {Nature Physics}\ }\textbf {\bibinfo {volume} {16}},\ \bibinfo {pages} {281} (\bibinfo {year} {2020})},\ \Eprint {https://arxiv.org/abs/1905.09408} {arXiv:1905.09408} \BibitemShut {NoStop}%
\bibitem [{\citenamefont {Gatto}\ \emph {et~al.}(2019)\citenamefont {Gatto}, \citenamefont {Facchi}, \citenamefont {Narducci},\ and\ \citenamefont {Tamma}}]{Gatto2019}%
  \BibitemOpen
  \bibfield  {author} {\bibinfo {author} {\bibfnamefont {D.}~\bibnamefont {Gatto}}, \bibinfo {author} {\bibfnamefont {P.}~\bibnamefont {Facchi}}, \bibinfo {author} {\bibfnamefont {F.~A.}\ \bibnamefont {Narducci}},\ and\ \bibinfo {author} {\bibfnamefont {V.}~\bibnamefont {Tamma}},\ }\href {https://doi.org/10.1103/PhysRevResearch.1.032024} {\bibfield  {journal} {\bibinfo  {journal} {Phys. Rev. Research}\ }\textbf {\bibinfo {volume} {1}},\ \bibinfo {pages} {032024(R)} (\bibinfo {year} {2019})}\BibitemShut {NoStop}%
\bibitem [{\citenamefont {Ge}\ \emph {et~al.}(2018)\citenamefont {Ge}, \citenamefont {Jacobs}, \citenamefont {Eldredge}, \citenamefont {Gorshkov},\ and\ \citenamefont {Foss-Feig}}]{Ge2018}%
  \BibitemOpen
  \bibfield  {author} {\bibinfo {author} {\bibfnamefont {W.}~\bibnamefont {Ge}}, \bibinfo {author} {\bibfnamefont {K.}~\bibnamefont {Jacobs}}, \bibinfo {author} {\bibfnamefont {Z.}~\bibnamefont {Eldredge}}, \bibinfo {author} {\bibfnamefont {A.~V.}\ \bibnamefont {Gorshkov}},\ and\ \bibinfo {author} {\bibfnamefont {M.}~\bibnamefont {Foss-Feig}},\ }\href {https://doi.org/10.1103/PhysRevLett.121.043604} {\bibfield  {journal} {\bibinfo  {journal} {Physical Review Letters}\ }\textbf {\bibinfo {volume} {121}},\ \bibinfo {pages} {043604} (\bibinfo {year} {2018})}\BibitemShut {NoStop}%
\bibitem [{\citenamefont {Tucker}\ \emph {et~al.}(1988)\citenamefont {Tucker}, \citenamefont {Eisenstein},\ and\ \citenamefont {Korotky}}]{Tucker1988}%
  \BibitemOpen
  \bibfield  {author} {\bibinfo {author} {\bibfnamefont {R.}~\bibnamefont {Tucker}}, \bibinfo {author} {\bibfnamefont {G.}~\bibnamefont {Eisenstein}},\ and\ \bibinfo {author} {\bibfnamefont {S.}~\bibnamefont {Korotky}},\ }\href {https://doi.org/10.1109/50.9991} {\bibfield  {journal} {\bibinfo  {journal} {Journal of Lightwave Technology}\ }\textbf {\bibinfo {volume} {6}},\ \bibinfo {pages} {1737} (\bibinfo {year} {1988})}\BibitemShut {NoStop}%
\bibitem [{\citenamefont {Redding}\ \emph {et~al.}(2020)\citenamefont {Redding}, \citenamefont {Murray}, \citenamefont {Donko}, \citenamefont {Beresna}, \citenamefont {Masoudi},\ and\ \citenamefont {Brambilla}}]{Redding2020}%
  \BibitemOpen
  \bibfield  {author} {\bibinfo {author} {\bibfnamefont {B.}~\bibnamefont {Redding}}, \bibinfo {author} {\bibfnamefont {M.~J.}\ \bibnamefont {Murray}}, \bibinfo {author} {\bibfnamefont {A.}~\bibnamefont {Donko}}, \bibinfo {author} {\bibfnamefont {M.}~\bibnamefont {Beresna}}, \bibinfo {author} {\bibfnamefont {A.}~\bibnamefont {Masoudi}},\ and\ \bibinfo {author} {\bibfnamefont {G.}~\bibnamefont {Brambilla}},\ }\href {https://doi.org/10.1364/OE.389212} {\bibfield  {journal} {\bibinfo  {journal} {Optics Express}\ }\textbf {\bibinfo {volume} {28}},\ \bibinfo {pages} {14638} (\bibinfo {year} {2020})}\BibitemShut {NoStop}%
\bibitem [{\citenamefont {Kirkendall}\ and\ \citenamefont {Dandridge}(2004)}]{Kirkendall2004}%
  \BibitemOpen
  \bibfield  {author} {\bibinfo {author} {\bibfnamefont {C.~K.}\ \bibnamefont {Kirkendall}}\ and\ \bibinfo {author} {\bibfnamefont {A.}~\bibnamefont {Dandridge}},\ }\href {https://doi.org/10.1088/0022-3727/37/18/R01} {\bibfield  {journal} {\bibinfo  {journal} {Journal of Physics D: Applied Physics}\ }\textbf {\bibinfo {volume} {37}},\ \bibinfo {pages} {R197} (\bibinfo {year} {2004})}\BibitemShut {NoStop}%
\bibitem [{\citenamefont {Lu}\ \emph {et~al.}(2019)\citenamefont {Lu}, \citenamefont {Lalam}, \citenamefont {Badar}, \citenamefont {Liu}, \citenamefont {Chorpening}, \citenamefont {Buric},\ and\ \citenamefont {Ohodnicki}}]{Lu2019}%
  \BibitemOpen
  \bibfield  {author} {\bibinfo {author} {\bibfnamefont {P.}~\bibnamefont {Lu}}, \bibinfo {author} {\bibfnamefont {N.}~\bibnamefont {Lalam}}, \bibinfo {author} {\bibfnamefont {M.}~\bibnamefont {Badar}}, \bibinfo {author} {\bibfnamefont {B.}~\bibnamefont {Liu}}, \bibinfo {author} {\bibfnamefont {B.~T.}\ \bibnamefont {Chorpening}}, \bibinfo {author} {\bibfnamefont {M.~P.}\ \bibnamefont {Buric}},\ and\ \bibinfo {author} {\bibfnamefont {P.~R.}\ \bibnamefont {Ohodnicki}},\ }\href {https://doi.org/10.1063/1.5113955} {\bibfield  {journal} {\bibinfo  {journal} {Applied Physics Reviews}\ }\textbf {\bibinfo {volume} {6}},\ \bibinfo {pages} {041302} (\bibinfo {year} {2019})}\BibitemShut {NoStop}%
\bibitem [{\citenamefont {Wu}\ \emph {et~al.}(2019)\citenamefont {Wu}, \citenamefont {Fan}, \citenamefont {Liu},\ and\ \citenamefont {He}}]{Wu2019}%
  \BibitemOpen
  \bibfield  {author} {\bibinfo {author} {\bibfnamefont {M.}~\bibnamefont {Wu}}, \bibinfo {author} {\bibfnamefont {X.}~\bibnamefont {Fan}}, \bibinfo {author} {\bibfnamefont {Q.}~\bibnamefont {Liu}},\ and\ \bibinfo {author} {\bibfnamefont {Z.}~\bibnamefont {He}},\ }\href {https://doi.org/10.1364/OL.44.005969} {\bibfield  {journal} {\bibinfo  {journal} {Optics Letters}\ }\textbf {\bibinfo {volume} {44}},\ \bibinfo {pages} {5969} (\bibinfo {year} {2019})}\BibitemShut {NoStop}%
\bibitem [{\citenamefont {Helstrom}(1969)}]{Helstrom1969}%
  \BibitemOpen
  \bibfield  {author} {\bibinfo {author} {\bibfnamefont {C.~W.}\ \bibnamefont {Helstrom}},\ }\href {https://doi.org/10.1007/BF01007479} {\bibfield  {journal} {\bibinfo  {journal} {Journal of Statistical Physics}\ }\textbf {\bibinfo {volume} {1}},\ \bibinfo {pages} {231} (\bibinfo {year} {1969})}\BibitemShut {NoStop}%
\bibitem [{\citenamefont {Ataman}(2019)}]{Ataman2019}%
  \BibitemOpen
  \bibfield  {author} {\bibinfo {author} {\bibfnamefont {S.}~\bibnamefont {Ataman}},\ }\href {https://doi.org/10.1103/PhysRevA.100.063821} {\bibfield  {journal} {\bibinfo  {journal} {Physical Review A}\ }\textbf {\bibinfo {volume} {100}},\ \bibinfo {pages} {063821} (\bibinfo {year} {2019})},\ \Eprint {https://arxiv.org/abs/1912.04018} {arXiv:1912.04018} \BibitemShut {NoStop}%
\bibitem [{\citenamefont {Lvovsky}(2015)}]{LvovskyPhotonicsch5}%
  \BibitemOpen
  \bibfield  {author} {\bibinfo {author} {\bibfnamefont {A.~I.}\ \bibnamefont {Lvovsky}},\ }\bibinfo {title} {Squeezed light},\ in\ \href {https://doi.org/https://doi.org/10.1002/9781119009719.ch5} {\emph {\bibinfo {booktitle} {Photonics}}}\ (\bibinfo  {publisher} {John Wiley and Sons, Ltd},\ \bibinfo {year} {2015})\ Chap.~\bibinfo {chapter} {5}, pp.\ \bibinfo {pages} {121--163}\BibitemShut {NoStop}%
\bibitem [{\citenamefont {Adesso}\ \emph {et~al.}(2014)\citenamefont {Adesso}, \citenamefont {Ragy},\ and\ \citenamefont {Lee}}]{Adesso2014}%
  \BibitemOpen
  \bibfield  {author} {\bibinfo {author} {\bibfnamefont {G.}~\bibnamefont {Adesso}}, \bibinfo {author} {\bibfnamefont {S.}~\bibnamefont {Ragy}},\ and\ \bibinfo {author} {\bibfnamefont {A.~R.}\ \bibnamefont {Lee}},\ }\href {https://doi.org/10.1142/S1230161214400010} {\bibfield  {journal} {\bibinfo  {journal} {Open Systems and Information Dynamics}\ }\textbf {\bibinfo {volume} {21}},\ \bibinfo {pages} {1440001} (\bibinfo {year} {2014})}\BibitemShut {NoStop}%
\bibitem [{\citenamefont {Serafini}(2017)}]{Serafini2017}%
  \BibitemOpen
  \bibfield  {author} {\bibinfo {author} {\bibfnamefont {A.}~\bibnamefont {Serafini}},\ }\href {https://doi.org/10.1201/9781315118727} {\emph {\bibinfo {title} {{Quantum Continuous Variables}}}}\ (\bibinfo  {publisher} {CRC Press},\ \bibinfo {address} {Boca Raton, FL : CRC Press, Taylor and Francis Group, [2017] |},\ \bibinfo {year} {2017})\BibitemShut {NoStop}%
\bibitem [{\citenamefont {Krueper}\ \emph {et~al.}(2024)\citenamefont {Krueper}, \citenamefont {Cohen},\ and\ \citenamefont {Gopinath}}]{github}%
  \BibitemOpen
  \bibfield  {author} {\bibinfo {author} {\bibfnamefont {G.}~\bibnamefont {Krueper}}, \bibinfo {author} {\bibfnamefont {L.}~\bibnamefont {Cohen}},\ and\ \bibinfo {author} {\bibfnamefont {J.~T.}\ \bibnamefont {Gopinath}},\ }\href {https://github.com/g-krueper/CascadedSensing} {\bibinfo {title} {{CascadedSensing}}},\ \bibinfo {howpublished} {\url{https://github.com/g-krueper/CascadedSensing}} (\bibinfo {year} {2024})\BibitemShut {NoStop}%
\bibitem [{\citenamefont {Mehmet}\ \emph {et~al.}(2010)\citenamefont {Mehmet}, \citenamefont {Eberle}, \citenamefont {Steinlechner}, \citenamefont {Vahlbruch},\ and\ \citenamefont {Schnabel}}]{Mehmet2010}%
  \BibitemOpen
  \bibfield  {author} {\bibinfo {author} {\bibfnamefont {M.}~\bibnamefont {Mehmet}}, \bibinfo {author} {\bibfnamefont {T.}~\bibnamefont {Eberle}}, \bibinfo {author} {\bibfnamefont {S.}~\bibnamefont {Steinlechner}}, \bibinfo {author} {\bibfnamefont {H.}~\bibnamefont {Vahlbruch}},\ and\ \bibinfo {author} {\bibfnamefont {R.}~\bibnamefont {Schnabel}},\ }\href {https://doi.org/10.1364/OL.35.001665} {\bibfield  {journal} {\bibinfo  {journal} {Optics Letters}\ }\textbf {\bibinfo {volume} {35}},\ \bibinfo {pages} {1665} (\bibinfo {year} {2010})}\BibitemShut {NoStop}%
\bibitem [{\citenamefont {Vernon}\ \emph {et~al.}(2019)\citenamefont {Vernon}, \citenamefont {Quesada}, \citenamefont {Liscidini}, \citenamefont {Morrison}, \citenamefont {Menotti}, \citenamefont {Tan},\ and\ \citenamefont {Sipe}}]{Vernon2019}%
  \BibitemOpen
  \bibfield  {author} {\bibinfo {author} {\bibfnamefont {Z.}~\bibnamefont {Vernon}}, \bibinfo {author} {\bibfnamefont {N.}~\bibnamefont {Quesada}}, \bibinfo {author} {\bibfnamefont {M.}~\bibnamefont {Liscidini}}, \bibinfo {author} {\bibfnamefont {B.}~\bibnamefont {Morrison}}, \bibinfo {author} {\bibfnamefont {M.}~\bibnamefont {Menotti}}, \bibinfo {author} {\bibfnamefont {K.}~\bibnamefont {Tan}},\ and\ \bibinfo {author} {\bibfnamefont {J.}~\bibnamefont {Sipe}},\ }\href {https://doi.org/10.1103/PhysRevApplied.12.064024} {\bibfield  {journal} {\bibinfo  {journal} {Physical Review Applied}\ }\textbf {\bibinfo {volume} {12}},\ \bibinfo {pages} {064024} (\bibinfo {year} {2019})},\ \Eprint {https://arxiv.org/abs/1807.00044} {arXiv:1807.00044} \BibitemShut {NoStop}%
\bibitem [{\citenamefont {Eberle}\ \emph {et~al.}(2011)\citenamefont {Eberle}, \citenamefont {H{\"{a}}ndchen}, \citenamefont {Duhme}, \citenamefont {Franz}, \citenamefont {Werner},\ and\ \citenamefont {Schnabel}}]{Eberle2011}%
  \BibitemOpen
  \bibfield  {author} {\bibinfo {author} {\bibfnamefont {T.}~\bibnamefont {Eberle}}, \bibinfo {author} {\bibfnamefont {V.}~\bibnamefont {H{\"{a}}ndchen}}, \bibinfo {author} {\bibfnamefont {J.}~\bibnamefont {Duhme}}, \bibinfo {author} {\bibfnamefont {T.}~\bibnamefont {Franz}}, \bibinfo {author} {\bibfnamefont {R.~F.}\ \bibnamefont {Werner}},\ and\ \bibinfo {author} {\bibfnamefont {R.}~\bibnamefont {Schnabel}},\ }\href {https://doi.org/10.1103/PhysRevA.83.052329} {\bibfield  {journal} {\bibinfo  {journal} {Physical Review A}\ }\textbf {\bibinfo {volume} {83}},\ \bibinfo {pages} {052329} (\bibinfo {year} {2011})},\ \Eprint {https://arxiv.org/abs/1103.1817} {arXiv:1103.1817} \BibitemShut {NoStop}%
\bibitem [{\citenamefont {Jarzyna}\ and\ \citenamefont {Demkowicz-Dobrza{\'{n}}ski}(2012)}]{Jarzyna2012}%
  \BibitemOpen
  \bibfield  {author} {\bibinfo {author} {\bibfnamefont {M.}~\bibnamefont {Jarzyna}}\ and\ \bibinfo {author} {\bibfnamefont {R.}~\bibnamefont {Demkowicz-Dobrza{\'{n}}ski}},\ }\href {https://doi.org/10.1103/PhysRevA.85.011801} {\bibfield  {journal} {\bibinfo  {journal} {Physical Review A}\ }\textbf {\bibinfo {volume} {85}},\ \bibinfo {pages} {011801(R)} (\bibinfo {year} {2012})},\ \Eprint {https://arxiv.org/abs/1108.3844} {arXiv:1108.3844} \BibitemShut {NoStop}%
\bibitem [{\citenamefont {Hecht}(2002)}]{hecht2002}%
  \BibitemOpen
  \bibfield  {author} {\bibinfo {author} {\bibfnamefont {E.}~\bibnamefont {Hecht}},\ }in\ \href@noop {} {\emph {\bibinfo {booktitle} {Optics}}}\ (\bibinfo  {publisher} {Addison-Wesley},\ \bibinfo {year} {2002})\ pp.\ \bibinfo {pages} {416--425}\BibitemShut {NoStop}%
\bibitem [{\citenamefont {Ono}\ and\ \citenamefont {Hofmann}(2010)}]{Ono2010}%
  \BibitemOpen
  \bibfield  {author} {\bibinfo {author} {\bibfnamefont {T.}~\bibnamefont {Ono}}\ and\ \bibinfo {author} {\bibfnamefont {H.~F.}\ \bibnamefont {Hofmann}},\ }\href {https://doi.org/10.1103/PhysRevA.81.033819} {\bibfield  {journal} {\bibinfo  {journal} {Physical Review A}\ }\textbf {\bibinfo {volume} {81}},\ \bibinfo {pages} {033819} (\bibinfo {year} {2010})},\ \Eprint {https://arxiv.org/abs/0910.3727} {arXiv:0910.3727} \BibitemShut {NoStop}%
\bibitem [{\citenamefont {{\v{S}}afr{\'{a}}nek}(2019)}]{Safranek2019}%
  \BibitemOpen
  \bibfield  {author} {\bibinfo {author} {\bibfnamefont {D.}~\bibnamefont {{\v{S}}afr{\'{a}}nek}},\ }\href {https://doi.org/10.1088/1751-8121/aaf068} {\bibfield  {journal} {\bibinfo  {journal} {Journal of Physics A: Mathematical and Theoretical}\ }\textbf {\bibinfo {volume} {52}},\ \bibinfo {pages} {035304} (\bibinfo {year} {2019})},\ \Eprint {https://arxiv.org/abs/1801.00299} {arXiv:1801.00299} \BibitemShut {NoStop}%
\bibitem [{\citenamefont {Storn}\ and\ \citenamefont {Price}(1997)}]{Storn1997}%
  \BibitemOpen
  \bibfield  {author} {\bibinfo {author} {\bibfnamefont {R.}~\bibnamefont {Storn}}\ and\ \bibinfo {author} {\bibfnamefont {K.}~\bibnamefont {Price}},\ }\href {https://doi.org/10.1023/A:1008202821328} {\bibfield  {journal} {\bibinfo  {journal} {Journal of Global Optimization}\ }\textbf {\bibinfo {volume} {11}},\ \bibinfo {pages} {341 } (\bibinfo {year} {1997})}\BibitemShut {NoStop}%
\bibitem [{\citenamefont {Williams}\ \emph {et~al.}(2012)\citenamefont {Williams}, \citenamefont {Jovanovic}, \citenamefont {Marshall}, \citenamefont {Smith}, \citenamefont {Steel},\ and\ \citenamefont {Withford}}]{Williams2012}%
  \BibitemOpen
  \bibfield  {author} {\bibinfo {author} {\bibfnamefont {R.~J.}\ \bibnamefont {Williams}}, \bibinfo {author} {\bibfnamefont {N.}~\bibnamefont {Jovanovic}}, \bibinfo {author} {\bibfnamefont {G.~D.}\ \bibnamefont {Marshall}}, \bibinfo {author} {\bibfnamefont {G.~N.}\ \bibnamefont {Smith}}, \bibinfo {author} {\bibfnamefont {M.~J.}\ \bibnamefont {Steel}},\ and\ \bibinfo {author} {\bibfnamefont {M.~J.}\ \bibnamefont {Withford}},\ }\href {https://doi.org/10.1364/OE.20.013451} {\bibfield  {journal} {\bibinfo  {journal} {Optics Express}\ }\textbf {\bibinfo {volume} {20}},\ \bibinfo {pages} {13451} (\bibinfo {year} {2012})}\BibitemShut {NoStop}%
\bibitem [{\citenamefont {Demkowicz-Dobrza{\'{n}}ski}\ \emph {et~al.}(2020)\citenamefont {Demkowicz-Dobrza{\'{n}}ski}, \citenamefont {G{\'{o}}recki},\ and\ \citenamefont {Guţă}}]{Demkowicz-Dobrzanski2020}%
  \BibitemOpen
  \bibfield  {author} {\bibinfo {author} {\bibfnamefont {R.}~\bibnamefont {Demkowicz-Dobrza{\'{n}}ski}}, \bibinfo {author} {\bibfnamefont {W.}~\bibnamefont {G{\'{o}}recki}},\ and\ \bibinfo {author} {\bibfnamefont {M.}~\bibnamefont {Guţă}},\ }\href {https://doi.org/10.1088/1751-8121/ab8ef3} {\bibfield  {journal} {\bibinfo  {journal} {Journal of Physics A: Mathematical and Theoretical}\ }\textbf {\bibinfo {volume} {53}},\ \bibinfo {pages} {363001} (\bibinfo {year} {2020})},\ \Eprint {https://arxiv.org/abs/2001.11742} {arXiv:2001.11742} \BibitemShut {NoStop}%
\bibitem [{\citenamefont {Kim}\ and\ \citenamefont {Song}(2016)}]{Kim2016}%
  \BibitemOpen
  \bibfield  {author} {\bibinfo {author} {\bibfnamefont {J.}~\bibnamefont {Kim}}\ and\ \bibinfo {author} {\bibfnamefont {Y.}~\bibnamefont {Song}},\ }\href {https://doi.org/10.1364/AOP.8.000465} {\bibfield  {journal} {\bibinfo  {journal} {Adv. Opt. Photon.}\ }\textbf {\bibinfo {volume} {8}},\ \bibinfo {pages} {465} (\bibinfo {year} {2016})}\BibitemShut {NoStop}%
\bibitem [{RFJ(2024)}]{RFJitter}%
  \BibitemOpen
  \href {https://www.berkeleynucleonics.com/model-745-oem-1-ps-delay-resolution} {\bibinfo {title} {Model 745-oem}} (\bibinfo {year} {2024})\BibitemShut {NoStop}%
\bibitem [{\citenamefont {Madsen}\ \emph {et~al.}(2022)\citenamefont {Madsen}, \citenamefont {Laudenbach}, \citenamefont {Askarani}, \citenamefont {Rortais}, \citenamefont {Vincent}, \citenamefont {Bulmer}, \citenamefont {Miatto}, \citenamefont {Neuhaus}, \citenamefont {Helt}, \citenamefont {Collins}, \citenamefont {Lita}, \citenamefont {Gerrits}, \citenamefont {Nam}, \citenamefont {Vaidya}, \citenamefont {Menotti}, \citenamefont {Dhand}, \citenamefont {Vernon}, \citenamefont {Quesada},\ and\ \citenamefont {Lavoie}}]{Madsen2022}%
  \BibitemOpen
  \bibfield  {author} {\bibinfo {author} {\bibfnamefont {L.~S.}\ \bibnamefont {Madsen}}, \bibinfo {author} {\bibfnamefont {F.}~\bibnamefont {Laudenbach}}, \bibinfo {author} {\bibfnamefont {M.~F.}\ \bibnamefont {Askarani}}, \bibinfo {author} {\bibfnamefont {F.}~\bibnamefont {Rortais}}, \bibinfo {author} {\bibfnamefont {T.}~\bibnamefont {Vincent}}, \bibinfo {author} {\bibfnamefont {J.~F.~F.}\ \bibnamefont {Bulmer}}, \bibinfo {author} {\bibfnamefont {F.~M.}\ \bibnamefont {Miatto}}, \bibinfo {author} {\bibfnamefont {L.}~\bibnamefont {Neuhaus}}, \bibinfo {author} {\bibfnamefont {L.~G.}\ \bibnamefont {Helt}}, \bibinfo {author} {\bibfnamefont {M.~J.}\ \bibnamefont {Collins}}, \bibinfo {author} {\bibfnamefont {A.~E.}\ \bibnamefont {Lita}}, \bibinfo {author} {\bibfnamefont {T.}~\bibnamefont {Gerrits}}, \bibinfo {author} {\bibfnamefont {S.~W.}\ \bibnamefont {Nam}}, \bibinfo {author} {\bibfnamefont {V.~D.}\ \bibnamefont {Vaidya}}, \bibinfo {author} {\bibfnamefont {M.}~\bibnamefont {Menotti}}, \bibinfo {author}
  {\bibfnamefont {I.}~\bibnamefont {Dhand}}, \bibinfo {author} {\bibfnamefont {Z.}~\bibnamefont {Vernon}}, \bibinfo {author} {\bibfnamefont {N.}~\bibnamefont {Quesada}},\ and\ \bibinfo {author} {\bibfnamefont {J.}~\bibnamefont {Lavoie}},\ }\href {https://doi.org/10.1038/s41586-022-04725-x} {\bibfield  {journal} {\bibinfo  {journal} {Nature}\ }\textbf {\bibinfo {volume} {606}},\ \bibinfo {pages} {75} (\bibinfo {year} {2022})}\BibitemShut {NoStop}%
\bibitem [{\citenamefont {Lundin}\ \emph {et~al.}(2011)\citenamefont {Lundin}, \citenamefont {Guan},\ and\ \citenamefont {Svanberg}}]{Lundin2011}%
  \BibitemOpen
  \bibfield  {author} {\bibinfo {author} {\bibfnamefont {P.}~\bibnamefont {Lundin}}, \bibinfo {author} {\bibfnamefont {Z.}~\bibnamefont {Guan}},\ and\ \bibinfo {author} {\bibfnamefont {S.}~\bibnamefont {Svanberg}},\ }\href {https://doi.org/10.1364/AO.50.000373} {\bibfield  {journal} {\bibinfo  {journal} {Applied Optics}\ }\textbf {\bibinfo {volume} {50}},\ \bibinfo {pages} {373} (\bibinfo {year} {2011})}\BibitemShut {NoStop}%
\bibitem [{\citenamefont {Xie}\ \emph {et~al.}(2009)\citenamefont {Xie}, \citenamefont {Chen},\ and\ \citenamefont {Ren}}]{Xie2009}%
  \BibitemOpen
  \bibfield  {author} {\bibinfo {author} {\bibfnamefont {F.}~\bibnamefont {Xie}}, \bibinfo {author} {\bibfnamefont {Z.}~\bibnamefont {Chen}},\ and\ \bibinfo {author} {\bibfnamefont {J.}~\bibnamefont {Ren}},\ }\href {https://doi.org/10.1016/j.measurement.2009.04.009} {\bibfield  {journal} {\bibinfo  {journal} {Measurement}\ }\textbf {\bibinfo {volume} {42}},\ \bibinfo {pages} {1335} (\bibinfo {year} {2009})}\BibitemShut {NoStop}%
\bibitem [{\citenamefont {Fritsch}\ and\ \citenamefont {Adamovsky}(1981)}]{Fritsch1981}%
  \BibitemOpen
  \bibfield  {author} {\bibinfo {author} {\bibfnamefont {K.}~\bibnamefont {Fritsch}}\ and\ \bibinfo {author} {\bibfnamefont {G.}~\bibnamefont {Adamovsky}},\ }\href {https://doi.org/10.1063/1.1136740} {\bibfield  {journal} {\bibinfo  {journal} {Review of Scientific Instruments}\ }\textbf {\bibinfo {volume} {52}},\ \bibinfo {pages} {996} (\bibinfo {year} {1981})}\BibitemShut {NoStop}%
\bibitem [{\citenamefont {Tan}\ \emph {et~al.}(2017)\citenamefont {Tan}, \citenamefont {Wang}, \citenamefont {Saraf},\ and\ \citenamefont {Lipa}}]{Tan2017}%
  \BibitemOpen
  \bibfield  {author} {\bibinfo {author} {\bibfnamefont {S.}~\bibnamefont {Tan}}, \bibinfo {author} {\bibfnamefont {S.}~\bibnamefont {Wang}}, \bibinfo {author} {\bibfnamefont {S.}~\bibnamefont {Saraf}},\ and\ \bibinfo {author} {\bibfnamefont {J.~A.}\ \bibnamefont {Lipa}},\ }\href {https://doi.org/10.1364/OE.25.003578} {\bibfield  {journal} {\bibinfo  {journal} {Optics Express}\ }\textbf {\bibinfo {volume} {25}},\ \bibinfo {pages} {3578} (\bibinfo {year} {2017})},\ \Eprint {https://arxiv.org/abs/1203.3914} {1203.3914} \BibitemShut {NoStop}%
\bibitem [{\citenamefont {Gibert}\ \emph {et~al.}(2015)\citenamefont {Gibert}, \citenamefont {Nofrarias}, \citenamefont {Karnesis}, \citenamefont {Gesa}, \citenamefont {Mart{\'{i}}n}, \citenamefont {Mateos}, \citenamefont {Lobo}, \citenamefont {Flatscher}, \citenamefont {Gerardi}, \citenamefont {Burkhardt}, \citenamefont {Gerndt}, \citenamefont {Robertson}, \citenamefont {Ward}, \citenamefont {McNamara}, \citenamefont {Guzm{\'{a}}n}, \citenamefont {Hewitson}, \citenamefont {Diepholz}, \citenamefont {Reiche}, \citenamefont {Heinzel},\ and\ \citenamefont {Danzmann}}]{Gibert2015}%
  \BibitemOpen
  \bibfield  {author} {\bibinfo {author} {\bibfnamefont {F.}~\bibnamefont {Gibert}}, \bibinfo {author} {\bibfnamefont {M.}~\bibnamefont {Nofrarias}}, \bibinfo {author} {\bibfnamefont {N.}~\bibnamefont {Karnesis}}, \bibinfo {author} {\bibfnamefont {L.}~\bibnamefont {Gesa}}, \bibinfo {author} {\bibfnamefont {V.}~\bibnamefont {Mart{\'{i}}n}}, \bibinfo {author} {\bibfnamefont {I.}~\bibnamefont {Mateos}}, \bibinfo {author} {\bibfnamefont {A.}~\bibnamefont {Lobo}}, \bibinfo {author} {\bibfnamefont {R.}~\bibnamefont {Flatscher}}, \bibinfo {author} {\bibfnamefont {D.}~\bibnamefont {Gerardi}}, \bibinfo {author} {\bibfnamefont {J.}~\bibnamefont {Burkhardt}}, \bibinfo {author} {\bibfnamefont {R.}~\bibnamefont {Gerndt}}, \bibinfo {author} {\bibfnamefont {D.~I.}\ \bibnamefont {Robertson}}, \bibinfo {author} {\bibfnamefont {H.}~\bibnamefont {Ward}}, \bibinfo {author} {\bibfnamefont {P.~W.}\ \bibnamefont {McNamara}}, \bibinfo {author} {\bibfnamefont {F.}~\bibnamefont {Guzm{\'{a}}n}}, \bibinfo {author} {\bibfnamefont
  {M.}~\bibnamefont {Hewitson}}, \bibinfo {author} {\bibfnamefont {I.}~\bibnamefont {Diepholz}}, \bibinfo {author} {\bibfnamefont {J.}~\bibnamefont {Reiche}}, \bibinfo {author} {\bibfnamefont {G.}~\bibnamefont {Heinzel}},\ and\ \bibinfo {author} {\bibfnamefont {K.}~\bibnamefont {Danzmann}},\ }\href {https://doi.org/10.1088/0264-9381/32/4/045014} {\bibfield  {journal} {\bibinfo  {journal} {Classical and Quantum Gravity}\ }\textbf {\bibinfo {volume} {32}},\ \bibinfo {pages} {045014} (\bibinfo {year} {2015})}\BibitemShut {NoStop}%
\bibitem [{\citenamefont {Sharma}\ \emph {et~al.}(2022)\citenamefont {Sharma}, \citenamefont {Sanders},\ and\ \citenamefont {Wilde}}]{Sharma2022}%
  \BibitemOpen
  \bibfield  {author} {\bibinfo {author} {\bibfnamefont {K.}~\bibnamefont {Sharma}}, \bibinfo {author} {\bibfnamefont {B.~C.}\ \bibnamefont {Sanders}},\ and\ \bibinfo {author} {\bibfnamefont {M.~M.}\ \bibnamefont {Wilde}},\ }\href {https://doi.org/10.1103/PhysRevResearch.4.023066} {\bibfield  {journal} {\bibinfo  {journal} {Physical Review Research}\ }\textbf {\bibinfo {volume} {4}},\ \bibinfo {pages} {023066} (\bibinfo {year} {2022})}\BibitemShut {NoStop}%
\bibitem [{\citenamefont {Oh}\ \emph {et~al.}(2019)\citenamefont {Oh}, \citenamefont {Lee}, \citenamefont {Rockstuhl}, \citenamefont {Jeong}, \citenamefont {Kim}, \citenamefont {Nha},\ and\ \citenamefont {Lee}}]{Oh2019}%
  \BibitemOpen
  \bibfield  {author} {\bibinfo {author} {\bibfnamefont {C.}~\bibnamefont {Oh}}, \bibinfo {author} {\bibfnamefont {C.}~\bibnamefont {Lee}}, \bibinfo {author} {\bibfnamefont {C.}~\bibnamefont {Rockstuhl}}, \bibinfo {author} {\bibfnamefont {H.}~\bibnamefont {Jeong}}, \bibinfo {author} {\bibfnamefont {J.}~\bibnamefont {Kim}}, \bibinfo {author} {\bibfnamefont {H.}~\bibnamefont {Nha}},\ and\ \bibinfo {author} {\bibfnamefont {S.-Y.}\ \bibnamefont {Lee}},\ }\href {https://doi.org/10.1038/s41534-019-0124-4} {\bibfield  {journal} {\bibinfo  {journal} {npj Quantum Information}\ }\textbf {\bibinfo {volume} {5}},\ \bibinfo {pages} {10} (\bibinfo {year} {2019})},\ \Eprint {https://arxiv.org/abs/1805.08495} {arXiv:1805.08495} \BibitemShut {NoStop}%
\bibitem [{\citenamefont {Pezz{\'{e}}}\ and\ \citenamefont {Smerzi}(2008)}]{Pezze2008}%
  \BibitemOpen
  \bibfield  {author} {\bibinfo {author} {\bibfnamefont {L.}~\bibnamefont {Pezz{\'{e}}}}\ and\ \bibinfo {author} {\bibfnamefont {A.}~\bibnamefont {Smerzi}},\ }\href {https://doi.org/10.1103/PhysRevLett.100.073601} {\bibfield  {journal} {\bibinfo  {journal} {Physical Review Letters}\ }\textbf {\bibinfo {volume} {100}},\ \bibinfo {pages} {073601} (\bibinfo {year} {2008})},\ \Eprint {https://arxiv.org/abs/0705.4631} {arXiv:0705.4631} \BibitemShut {NoStop}%
\end{thebibliography}%

\end{document}